\documentclass{aa}  

\usepackage{graphicx}
\usepackage{txfonts}
\usepackage{subcaption}         
\usepackage{lscape}             
\usepackage{placeins}           
                                
\usepackage{xspace}

\newcommand{\ee}{$e^\pm$}

\newcommand{\sax}{{\it BeppoSAX}\xspace}

\newcommand{\batse}{{\it BATSE}\xspace}
\newcommand{\hete}{{\it HETE-2}\xspace}

\newcommand{\swift}{{\it Swift}\xspace}

\newcommand{\chandra}{{\it Chandra}\xspace}

\newcommand{\ginga}{{\it Ginga}\xspace}

\newcommand{\srobs}{\rm SR_{obs}}
\newcommand{\sr}{\rm SR}

\newcommand{\kev}{\rm keV~}

\newcommand{\fluence}{\rm erg\ cm^{-2}}
\newcommand{\cnu}{\chi^2_\nu}

\newcommand{\epc}{E_{\it p}}
\newcommand{\eo}{E_{\it o}}
\newcommand{\cq}{\chi^2}

\def\be{\begin{equation}}
\def\ee{\end{equation}}

\def\ltsima{$\; \buildrel < \over \sim \;$}
\def\lsim{\lower.5ex\hbox{\ltsima}}
\def\gtsima{$\;\buildrel > \over \sim \;$}
\def\gsim{\lower.5ex\hbox{\gtsima}}
\usepackage{xcolor}

\titlerunning{Properties of XRF versus GRBs from \sax}

\begin{document}

\title{Comparative properties of X-Ray Flashes and Gamma-Ray Bursts from BeppoSAX observations of Fast X-ray Transients}

\author{L. Piro
          \inst{1}
          \and
          G. Gianfagna
          \inst{1}
          \and
          J.J.M. in 't Zand \inst{2}
          \and
          B. Gendre\inst{3,4}
          \and
          C. Guidorzi \inst{5,6}
          \and
          L. Amati \inst{6}
          \and
          F. Frontera \inst{5,6}
          \and E. Kuulkers \inst{7}
}
\institute{INAF -- Istituto di Astrofisica e Planetologia Spaziali, via Fosso del Cavaliere 100, I-00133 Rome, Italy\\ 
\email{luigi.piro@inaf.it}
\and
SRON Space Research Organization Netherlands, Niels Bohrweg 4, 2333 CA Leiden, the Netherlands
\and
OzGrav ARC Centre of Excellence, The University of Western Australia, 35 Stirling Highway, 6009 Crawley, WA, Australia
\and
College of Science and Mathematics, University of the Virgin Islands, 2 John Brewer’s Bay, St Thomas, VI 00802, USA
\and
Dept. of Physics and Earth Science, University of Ferrara, 44122 Ferrara, Italy
\and
INAF-OAS Bologna, Via P. Gobetti 101, 40129, Bologna, Italy
\and
ESTEC, ESA, Keplerlaan 1, 2201 AZ Noordwijk, the Netherlands
}

\abstract{We present the homogeneous and complete sample of 96 bona-fide Gamma Ray Bursts (GRBs)
detected by the Wide Field Cameras (2-28 keV) of \textit{BeppoSAX}.
We complement these data with the Gamma-Ray Burst Monitor (40-700 keV) simultaneous observations. We derive the spectral and
temporal properties of the prompt emission,  and assess the properties
of the soft population of GRB, namely X-ray flashes (XRFs) in comparison with normal GRBs. On the
basis of the spectral shape we find that 36 events are XRFs, 40 X-ray rich events (XRR), and 20 normal GRBs. We analyze the distribution of the spectral parameters of
the Band function, $\alpha$, $\beta$ and $\epc$, finding that the spectral indexes of the three classes are broadly
similar. On the contrary the peak energy is the parameter driving the spectra shape, from 8.5 keV for XRF keV to 83 keV for GRBs. The  duration ($T_{90}$) in the X-ray range is similar
in  the three classes, clustering around 70 s. Likewise, a similar duration  of 25 s is observed in the gamma-ray  range.
For the 67  events that are detected in both instruments we find that 9 events exhibit a soft X-ray precursor taking place from 14 to 105 s before the onset of the gamma-ray burst. 
About 90\% of the events that were identified in real time 
exhibit an X-ray afterglow, with a similar fraction for the three classes. In the optical and radio the the corresponding fractions are 35\% and 33\%. 
All the similarities in the spectrum,  duration and afterglow properties suggest  common progenitors for the three classes, where the differences are likely a combination of the effect of different baryon loading, energy, structure and orientation of the jet with respect to the observer. A comparison  with \textit{Einstein Probe} shows that the latter, thanks to its sensitivity, reaches out to a population of fainter and more numerous events, whose presence was firstly hinted at by the unique very low luminosity \textit{BeppoSAX} GRB980425. 
}

\keywords{X-ray: general-Gamma-ray: burst}

\maketitle

\section{Introduction}

Fast X-Ray Transients (FXTs) were observed in the past by several
satellites \citep{arefiev}.   The origin of  events with durations larger than thousands of seconds was
attributed to  flare stars and  RS CVn systems \citep{gotthelf}, whereas shorter events were associated with GRBs by subsequent observations.
In \ginga data from 1987 to 1991, \citet{Strohmayer1998} identified a fraction of GRBs (36\%) with very soft spectra, but similar features to canonical GRBs observed by BATSE (Preece et al. 2000; Kaneko et al. 2006).
However, only with the launch of \sax  a systematic study of these soft events,  named X-ray Flashes (XRF), started.
\sax was an Italian-Dutch mission that operated between 1996 to 2002 \citep{Boella}. It had on board two Wide Field Camera (WFC, \citealt{Jager1997}) modules observing in the range 2-28 keV, and an open-sky Gamma-Ray Burst Monitor (GRBM, \citealt{Feroci_1997}) for the band 40-700 keV. XRFs were initially distinguished by GRBs by the lack of emission in the \sax GRBM,
while the X-ray emission in the WFCs was similar to normal GRBs \citep{heise}.

In the \sax observations, XRFs show an isotropic distribution on the sky and a duration
between few seconds and $\sim 10^{3}$ seconds, like long GRBs
\citep{heise}. The first identification of host galaxies,
confirming the cosmological origin of XRF, was obtained by \citet{Bloom2003}. Indeed their distribution in distance ($<z>\approx 1$) derived from \sax and \hete samples \citep{DAlessio2006, Sakamoto2005}, is similar to that of long GRB observed by the same satellites. A similar average
redshift $<z_{XRF}> = 1.40$ was derived by \citet{Gendre_2007} from a first sample of 9 XRFs observed by the Neil Gehrels Swift Observatory (\swift). Also in larger samples of \swift GRBs \citep{Sakamoto2008, Bi2018}, the mean redshift is $\approx 1.5$, similar to long GRBs.

With the increased sample of events catered for by \sax, \hete, and \swift, it appeared that the distribution of spectral hardness \citep{Kippen2003, Sakamoto2005, Sakamoto2008, Pelageon2008, Xu_2025} of XRF and GRB is not sharply divided but indeed is smoothly connected throughout the softer to the hardest events by
an intermediate class, named X-Ray Rich (XRR) Gamma Ray Bursts  \citep{barr}.

\citet{Kippen2003} analyzed a sample of 9 XRFs observed by \sax also using untriggered \batse data and found that the photon index $\alpha$ and $\beta$
of the Band function of the XRFs are similar to those of the GRBs,
instead of $\epc$ (spectral peak energy), whose value is less than 10 \kev for most XRFs.
This result has been confirmed by \citet{Sakamoto2005} with the
analysis of 42 XRRs/XRFs observed by \hete.

Discovering soft X-ray transients has been quite difficult after the end of the \sax mission, due to instruments mainly observing the sky in the hard X-rays and $\gamma-$rays. However, from 2024, the number of soft FXTs is growing thanks to \textit{Einstein Probe} (EP) mission. EP-WXT has been discovering since 2024 hundreds of these sources \citep{Aryan2025, Wu2025, Zhang2025}.

Several theories have been proposed to explain the origin of XRFs:
high redshift GRBs \citep{heise}, off-axis viewed GRBs with a uniform
jet \citep{yamaz02,yamaz03,yamazaki04}, with the Universal
Power-law-shaped jet \citep{lamb05},  with a Gaussian jet
\citep{zhang03}, with a ring shaped jet \citep{eichler} and with a multi sub-jets \citep{Toma_2005},  a variable jet opening-angle \citep{lamb05}, dirty  fireballs \citep{dermer}, a  photosphere dominated emission
\citep{ram} and  off-axis cannonballs \citep{dar}.
\citet{DAlessio2006} analyzed a sample of \sax and \hete bursts and test off-axis uniform and stuctured jet theories as origin of XRFs. They find that, while the uniform off-axis jet can be ruled out, a structured jet could explain XRF afterglow and prompt emission only if the burst is observed slightly outside the narrow central core but still within the outer envelope.

There are several studies of prompt and afterglow properties of individual XRRs/XRFs \citep{Frontera_2000, Crew2003, Amati2004, Sakamoto2006, Schady2006, stratta2007, Arimoto2007, Galli2006, Guidorzi2009}.
XRF 011030 was observed by \sax and has one of the longest durations in the WFC, about 1500 s 
\citep{Galli2006}. For this burst, it is possible that the WFC also detected the onset of the afterglow emission, observed also by \chandra at late times. \citet{Galli2006} find that the broadband afterglow data, including optical and radio measurements, are consistent either with a fireball expanding in a wind environment or with a jetted fireball in an ISM.

XRF 060218 is a nearby (36 Mpc) event \citep{Campana2006}. This is the first XRF to be associated with a Supernova (SN2006aj, \citealt{Pian2006}). Its soft prompt emission (with a spectral peak energy of about 5 keV) lasted 1200 s. Early X-ray and optical emission are dominated by a SN shock breakout. At late times, instead, the light curve is typical of GRB afterglows. This classical afterglow can be naturally accounted for by a shock driven into the wind by a mildly relativistic shell \citep{Soderberg2006}. 

XRF 020903 was detected by \hete, with a prompt spectral peak energy of 3.3 keV. Its multi-band afterglow light curve shows an achromatic brightening around 0.7 days. \citet{Urata2015} explain it with an off-axis jet model with a large observing angle (twice the jet opening half-angle). Also another XRF, 080330 (discovered by \swift, with a peak energy lower than 88 keV, \citealt{Guidorzi2009}), can be described with an off-axis model. 

EP240801 \citep{Jiang2025} was observed by EP, with a duration of 148 s in X-rays by EP WXT and 22 s in gamma-rays by \textit{Fermi-GBM}. Its spectrum has a peak energy of about 15 keV. 
Different models have been proposed to explain this XRF. 
A structured jet viewed off-axis, with two components: a narrow, high-Lorentz factor jet core (typical of classical GRBs) plus a wider, softer component. Another possibility is the presence of a prolonged central engine activity, that injects energy into the external shock, softening the overall spectrum.

So far  studies of  spectral properties of XRF  have been limited to individual or small samples of events. We previously presented a complete sample of FXT events detected with the \sax WFCs \citep[][hereafter paper I]{Zand2026}. In the present paper, we present a comparative analysis of all XRFs, XRRs and GRBs in this sample, focusing on their spectral properties and information related to their afterglows.

The paper is organized as follows. In Section \S~\ref{sec:catalogo} we present the analysed dataset, and in \S~\ref{sec:spectral_class} we describe the spectral fitting procedure. In \S~\ref{sec:properties_prompt}, and \ref{sec:properties_afterglow} we describe the prompt and afterglow properties respectively. Finally, in \S~ \ref{sec:discussion} we discuss our results, and in \S~\ref{sec:conclusioni} we present the conclusions.

\section{Observations and analysis}
\label{sec:catalogo} 

In paper I 100 FXT events detected by the WFCs are associated to GRBs. From these, we have excluded 4 events (970111b, 980306, 981217, 020410), as they have no full coverage by the GRBM, for a total of 96 events. 29 events are not detected in the GRBM.
For the full catalogue  of GRBM detected events see \citet{Frontera_2000, Guidorzi2011}.

For all these events we have produced the light curves of the GRBM, available with 1 sec resolution in two energy bins, 40-700 keV and $> 100$ keV, corresponding to the side of the instrument co-aligned with the corresponding WFC \citep{Feroci_1997}. To estimate the background counts in the GRBM light curves we perform either a linear or parabolic fit of the light curve in time intervals before and after the WFC burst. Each interval is equal to $2T_{90}$ of the WFC.
In Table \ref{tab1} we report the date, durations ($T_{50}$ and $T_{90}$) and average count rates for the two instruments. We estimated the average count rates of the WFC and GRBM in the same exposure time, corresponding to the one from the WFC. This allows us to estimate and compare the fluences in WFC and GRBM in the same time interval.
We note that, for this reason, some faint events that are detected in the GRBM (\citealt{Frontera_2000}, paper I) but have a long duration in the WFC, can result in a lower average count rate or non detection.
All events detected in the GRBM have a duration larger than 2s, i.e. they are long GRBs.
We have finally produced the overall spectrum of each event, including the WFC and the
spectral bins of the GRBM. The spectral data of the GRBM have been accumulated in the same
time interval where the WFC light curve is above the background. Inspection of the light
curves showed, in some cases, that a  GRBM signal starting  before the (WFC) trigger time.
This was expected, because of the well known hard to soft evolution \citep{Yoshida1989, Band1997, Piro1998}, and taken into account
by adding this time slice.

\begin{figure}
\centering
 \includegraphics[width=0.45\textwidth]{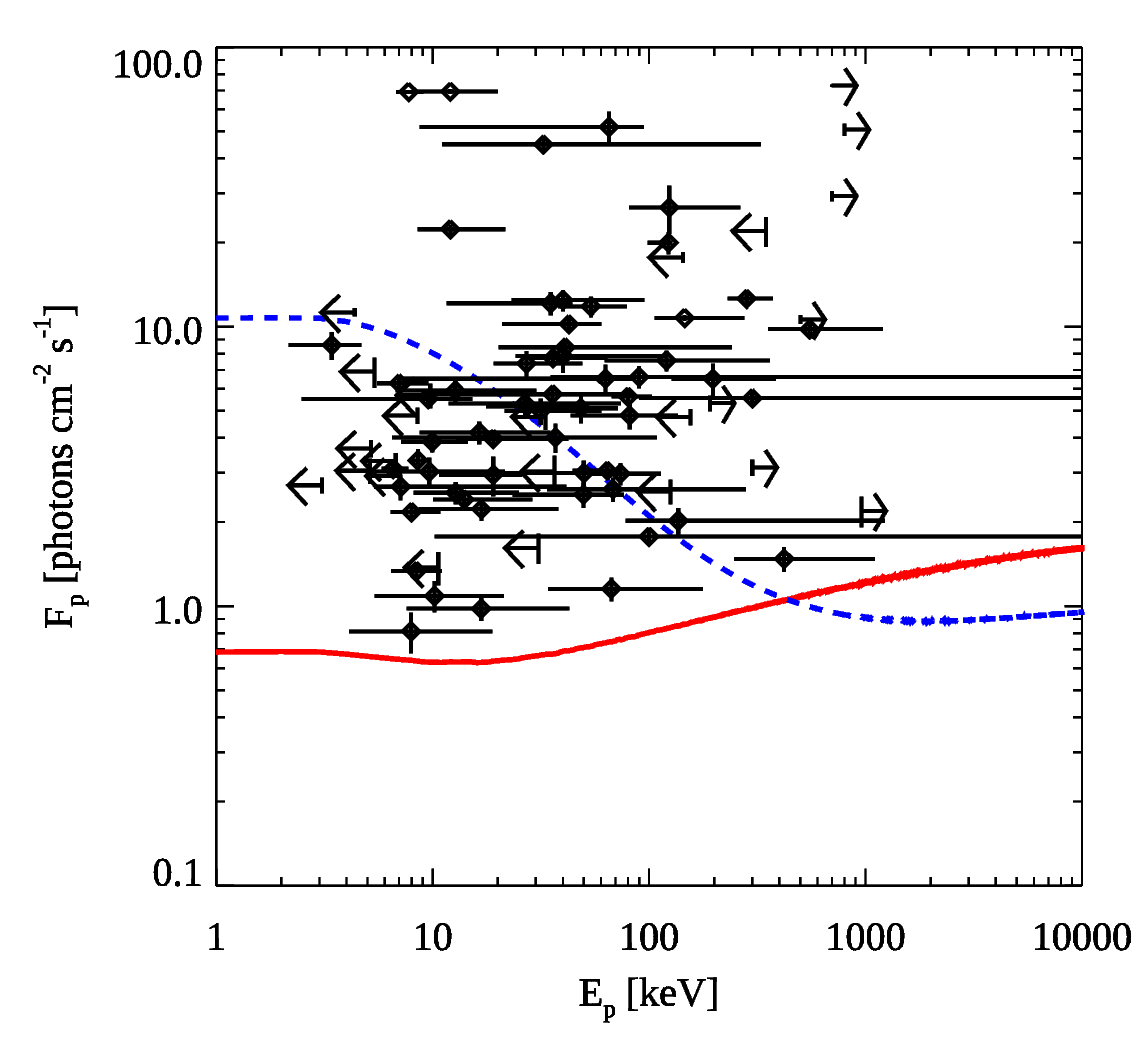}
  \caption{Sensitivity of the GBRM (blue dashed line) and WFC (red solid line) for 1 s integration time as a function of the peak flux (2-10000 keV) and peak energy $\epc$ (derived in Section \ref{sec:spectralparam}). The results on our sample are plotted (data points, upper and lower limits).  }
\label{fig:sensitivity}
\end{figure}

It is important to point out that the sample of GRBs selected by the WFC (2-28 keV) is not
significantly biased against hard GRBs. To demonstrate this point, we have reported in
Fig. \ref{fig:sensitivity} the sensitivity of the WFC and of the GRBM 
 as a function of the peak energy $\epc$ derived from Band model fitting (described in Section \ref{sec:spectralparam}) and of the peak flux in the range 2-10000 keV.
 Note as the WFC sensitivity is relatively
independent of the peak energy and 
remains comparable to that of the GRBM even for $\epc$
above 100 keV, while being significantly better below 100 keV. Thus even
the harder GRBs are above the detection threshold of the WFC. 
This figure outlines another key point. It shows that relatively bright X-ray flashes (e.g. $\epc$ less than about 30 keV) can be  detected in the GRBM, while fainter events are not. On the contrary, a detection in a gamma-ray experiment does not classify the event as a GRB. This underlines the general point that a classification of XRF vs GRB based solely on the non detection in a gamma-ray experiment is not sufficient.   It requires a proper evaluation of the lower limit on the softness ratio.

\section{Spectral classification}
\label{sec:spectral_class}

\begin{figure}
 \includegraphics[width=0.5\textwidth]{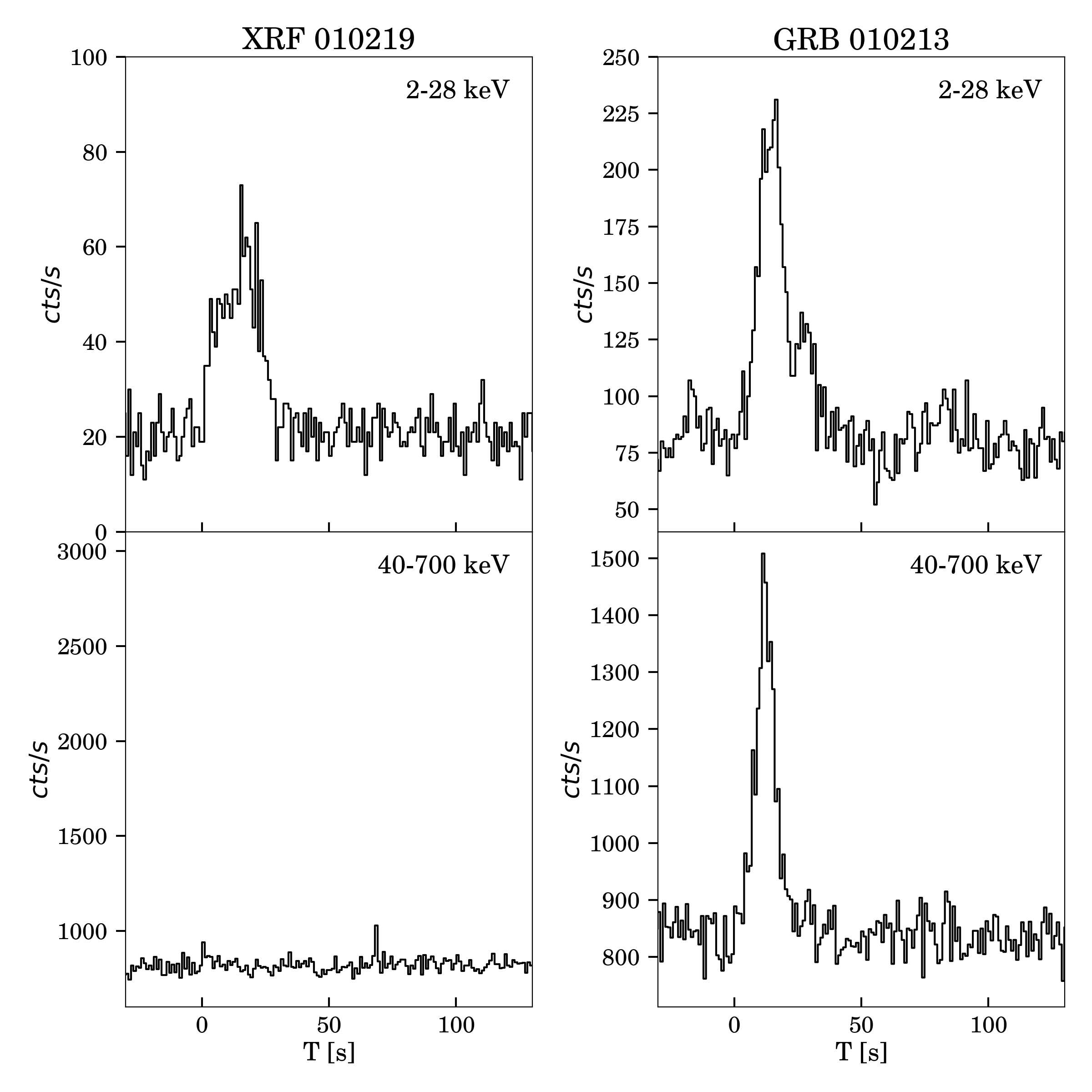}
\caption{Example of a GRB  
and an XRF 
light curves. }
\label{fig:GRB_XRF_lc}
\end{figure}

\begin{figure}
 \includegraphics[width=0.5\textwidth]{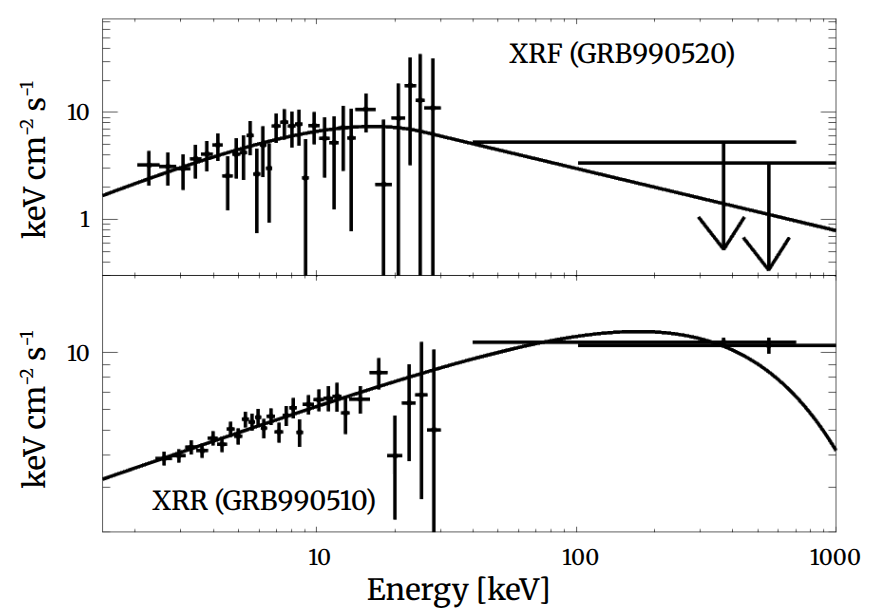}
\caption{Example of a XRR (990510) and an XRF (990520)  $\nu F_{\nu}$ spectra. The solid line represents the fitted Band model (see Table \ref{tab:spectra}). Note that the values of the Band function slopes $\alpha$ and $\beta$ represented in the plot have to be increased by 2 with respect to Table 
\ref{tab:spectra}.}
\label{fig:GRB_XRF_spec}
\end{figure}

We analyzed the spectral properties of our sample by using  the softness ratio as main
parameter.  We have first derived a preliminary classification adopting the {\it observed}
softness ratio $\srobs$, given by the ratio of the WFC (2-28 keV) and GRBM (40-700 keV) count
rates (Table \ref{tab1} and Fig.\ref{fig:SR_obs}). We have then calibrated $\srobs$ with the spectral softness ratio
SR defined by the ratio of fluences in the 2-30 keV and 30-400 keV ranges \citep{Sakamoto2005}.  We
have adopted a Band spectral model with $\alpha=-1$ and $\beta=-2.2$ and varying $\epc$. We
define the following criteria of classification. GRBs: $\srobs \le 0.53\times 10^{-2}$, corresponding to $\epc>100$ keV; XRR:
$0.53\times 10^{-2} \le \srobs \le 2.75\times 10^{-2}$, corresponding to $30 \rm\ keV <\epc<100 \rm\ keV$ and XRF: $\srobs \ge 2.75\times 10^{-2}$, with $\epc<30$ keV. Out of a total sample of 96 bursts, 23 are classified as GRBs, 37 XRRs and 36 as XRFs.
In Fig. \ref{fig:GRB_XRF_lc} and Fig. \ref{fig:GRB_XRF_spec} we show a comparison of the light curves and spectra of a GRB vs XRF. 
In order to derive an instrument independent softness ratio, we have refined our procedure by
fitting each spectrum with a Band model
\begin{equation}
N(E) =
\begin{cases}
A \left(\frac{E}{100\,\mathrm{keV}}\right)^{\alpha}
\exp\left(-\frac{E}{E_0}\right) & E < (\alpha - \beta)E_0 \\
A \left(\frac{(\alpha - \beta)E_0}{100\,\mathrm{keV}}\right)^{\alpha - \beta}
\exp(\beta - \alpha)
\left(\frac{E}{100\,\mathrm{keV}}\right)^{\beta} & E \ge (\alpha - \beta)E_0
\end{cases}
\end{equation}
where $\alpha$ ($\beta)$ is the low (high) spectral photon
index, $\epc$ is the peak energy in keV, estimated as $\epc =(2+\alpha)\times \eo$, $\eo$ is the break energy, and $A$ is the normalization in $\rm photons \ cm^{-2} s^{-1} keV^{-1}$. In all spectral fits, we accounted for the Galactic column density  along the line of sight (see paper I). 
Notwithstanding the broad energy range covered by the combination of \sax WFC and GRBM,
from 2 to 700 keV, we found that in a significant fraction of events, one or more of the
spectral parameters were not constrained. This is  due to one or a combination of the
following effects: small signal-to-noise ratio, the high energy range being covered by only two (GRBM) spectral
channels, peak energy being outside the energy range covered by the instruments. To cope with these problems, we have adopted the following procedure. All the spectra have at first been fitted
with a simple power law. Best fit slope $\Gamma_{PL}$ and $\cnu$ are reported in
Table \ref{tab:spectra}. We have then run a fit with the Band model and registered the
improvement in $\chi^2$. For those spectra where the improvement is not significant ($\Delta
\chi^2 \le 9$) we find that Band spectral parameters are poorly constrained. We therefore had
to chose which of either $\alpha$ or $\beta$ had to be fixed. We chose to fix $\alpha=-1$
when $\srobs \ge 1\times10^{-2}$ and $\beta=-2.2$ in the opposite condition. These values are taken from $\alpha$ and $\beta$ usually found in the literature \citep[for example][]{Sakamoto2005, Sakamoto2008}
In practice, the peak energy is either below or above the instrumental energy range, and this procedure forces
the fit to derive an upper limit on $\epc$ for soft events and a lower limit on $\epc$ for
hard events.

\begin{figure}
 \includegraphics[width=0.5\textwidth]{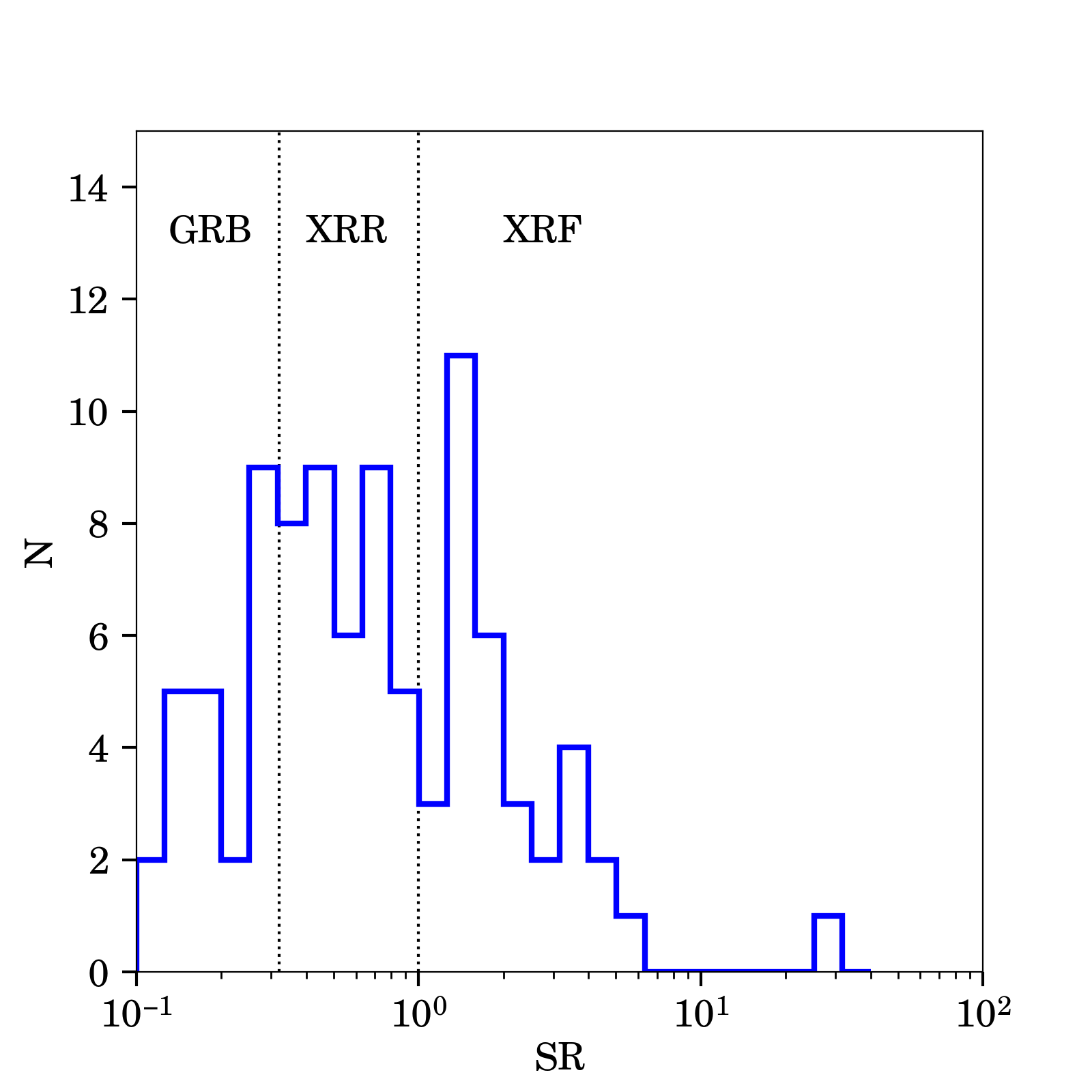}
\caption{The distribution of the softness ratio,  $SR=S(2,30)/S(30,400)$.  GRBs:  $\sr \le 0.32$, XRRs: $ 0.32 \le \sr \le 1$ and  XRFs: $ \sr \ge 1$.}
\label{fig:HR}
\end{figure}

The spectral parameters, the fluence and softness spectral ratio $\sr$ derived from the
best fit spectrum are reported in Table \ref{tab:spectra}. In Fig. \ref{fig:HR} we show the
distribution of $\sr$. Following \citet{Sakamoto2005} we finally classify the events as follows: GRB: $ \sr
\le 0.32$, XRR: $ 0.32 \le \sr \le 1$ and XRF: $ \sr \ge 1$. Out of a total sample of 96
bursts, we find 20 GRBs, 36 XRFs and 40 XRRs (see Table \ref{table:numbers}), comfortingly similar to the numbers derived
adopting the observed ratios. Minor differences are due to events at the boundaries of the classes.

\begin{table}
\caption{The sample of \sax GRBs.} 
\label{table:numbers}     
\centering                         
\begin{tabular}{c c c c}      
\hline\hline               
 & GRB & XRR & XRF \\   
\hline                   
Total & 20 & 40 & 36 \\
\hline
With GRBM & 19 & 34 & 14 \\
Precursors & 1 & 5 & 3 \\
\hline
Real time alert & 17 & 22 & 14 \\
X-ray follow-up & 15 & 13 & 9 \\
X-ray afterglow & 13 & 12 & 7 \\
Optical follow-up & 18 & 20 & 14 \\
Optical afterglow & 6 & 9 & 4 \\
Radio follow-up & 15 & 12 & 7 \\
Radio afterglow & 5 & 5 & 2 \\
\hline                                   
\end{tabular}
\tablefoot{
    The number of GRBs, XRRs and XRFs in the sample is written in the 1st row. In the 2nd and 3rd row there are the numbers of events (divided by class) with a GRBM counterpart, and with a soft X-ray precursor (see Section \ref{sec:precursors}). In the 4th row the number of bursts with a real time alert is reported. In the 5th and 6th, the number of bursts with an X-ray follow up and an X-ray detection of the afterglow are included. The same for the 7th-8th rows for optical, and 9th-10th for radio observations. } 
\end{table}

\section{Properties of the prompt emission}
\label{sec:properties_prompt}
\subsection{Spectra}
\label{sec:spectralparam}

\subsubsection{Spectral shape}
\label{subsec:spec_shape}

As described in the previous section, spectra were individually fit with a power law and then
with the Band model. Spectra were also tested for consistency with a black body model (see Appendix \ref{App:spectra} for some examples of spectral fits). We
have then compared the observed distribution of $\cq$  with that expected  assuming that the
model is true.
Out of 96 spectra, we expect 5 with  $\cnu>1.5$ for 27 degree of freedom ($\nu$). In the case of black body 79 spectra display
$\cq$ above this threshold, that confirms that this model is not adequate, being too narrow to account for the spectral extension
displayed by the data. For the power law the situation improves
but there are still 36 spectra above the limit. On the contrary, the
Band model is nicely consistent with the expected distribution,
with 5 spectra with $\cnu>1.5$.

We have further considered those spectra where the black body model appears to be consistent
with the data to verify if,  in these few cases,  this model provides a better fit than the
power law or the Band model. We find that there is no such case, actually in 13 cases the
black body model is significantly ($\Delta \cq>9$) worse, while for the remaining 5 cases
there is no significant difference.
We remark that we have used both WFC and GRBM spectra while in paper I only the data of WFC were fitted. This explains why in a few cases the simple power law spectral slopes derived here are steeper, being constrained by the higher energy data.

\subsubsection{Distribution of spectral parameters}
\begin{figure}
 \includegraphics[width=0.5\textwidth]{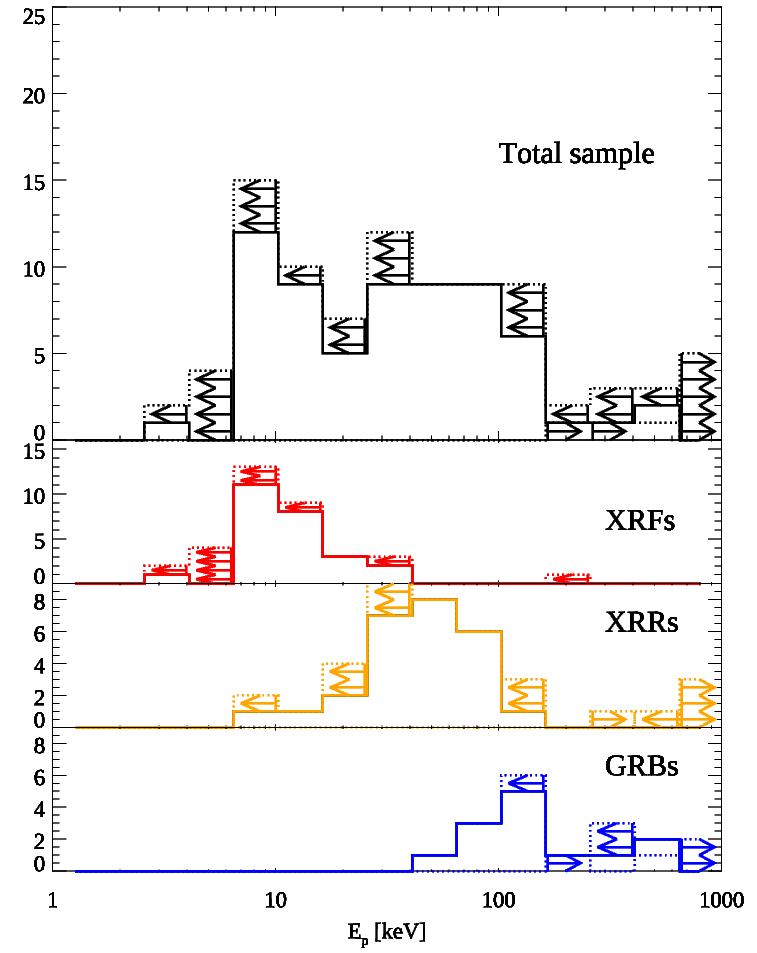}
\caption{Distribution of $\epc$ for, from the top, the full sample (96 bursts), XRF, XRR and GRB. Upper and lower limits are represented as arrows.}
 \label{fig:Ep}
\end{figure}

We studied and compared the prompt emission spectral parameters of XRRs/XRFs and  GRBs.  The distribution  of $\epc$, presented in Fig. \ref{fig:Ep} is very broad, spanning two orders of magnitude, from a few keV up to 1 MeV. This distribution (like $\alpha$ and $\beta$, described later) is in reality the result of the convolution of the intrinsic $\epc$ distribution (the so called parent distribution) with the measurement error. Under the assumption that both are Gaussian, it is possible to deconvolve the two distributions \citep[see][for the full procedure]{DePasquale2006}.  The intrinsic width of the parent distribution of $\epc$ (see Table \ref{tab:spectralaverage}) is not consistent with zero. Indeed, the corresponding distribution for each subclass (Fig. \ref{fig:Ep})  shows that for XRF  $<\epc^{XRF}>=8.5$ keV, a factor of about 10 times lower than that observed in GRB, $<\epc^{GRB}>=83$ keV. 

\begin{figure}
 \includegraphics[width=0.5\textwidth]{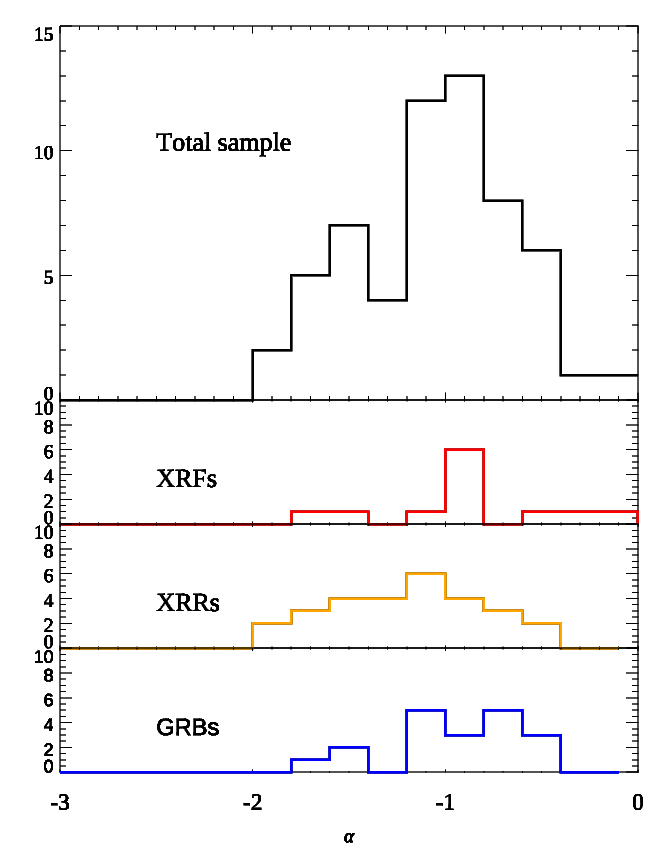}
\caption{Distribution of $\alpha$ for, from the top, all the sample (64 events with $\alpha$ as free parameter), XRF, XRR and GRB. No upper or lower limits are represented as none was found in the fits.}
 \label{fig:alpha}
\end{figure}

On the contrary, the intrinsic distribution of $\alpha$ is narrow. In fact its width is consistent with zero in the case of XRFs and XRRs, and is only marginally greater than zero for GRBs and for the full population, (see $\sigma_{\alpha}$ in Table \ref{tab:spectralaverage}). Moreover, there is no statistically significant difference in $\alpha$ for XRF, XRR, and GRB (see also Fig. \ref{fig:alpha}).

\begin{figure}
 \includegraphics[width=0.5\textwidth]{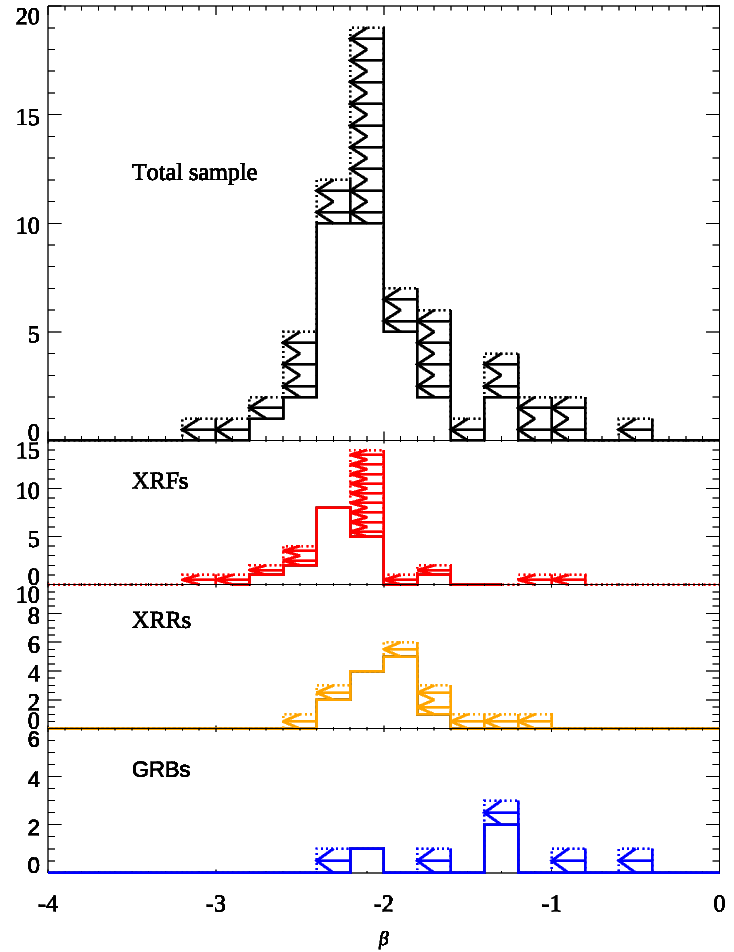}
\caption{Distribution of $\beta$ for, from the top, all the sample (67 events with $\beta$ as free parameter), XRF, XRR and GRB.  Lower limits are represented as arrows.}
 \label{fig:beta}
 \end{figure}

\begin{figure}
 \includegraphics[width=0.5\textwidth]{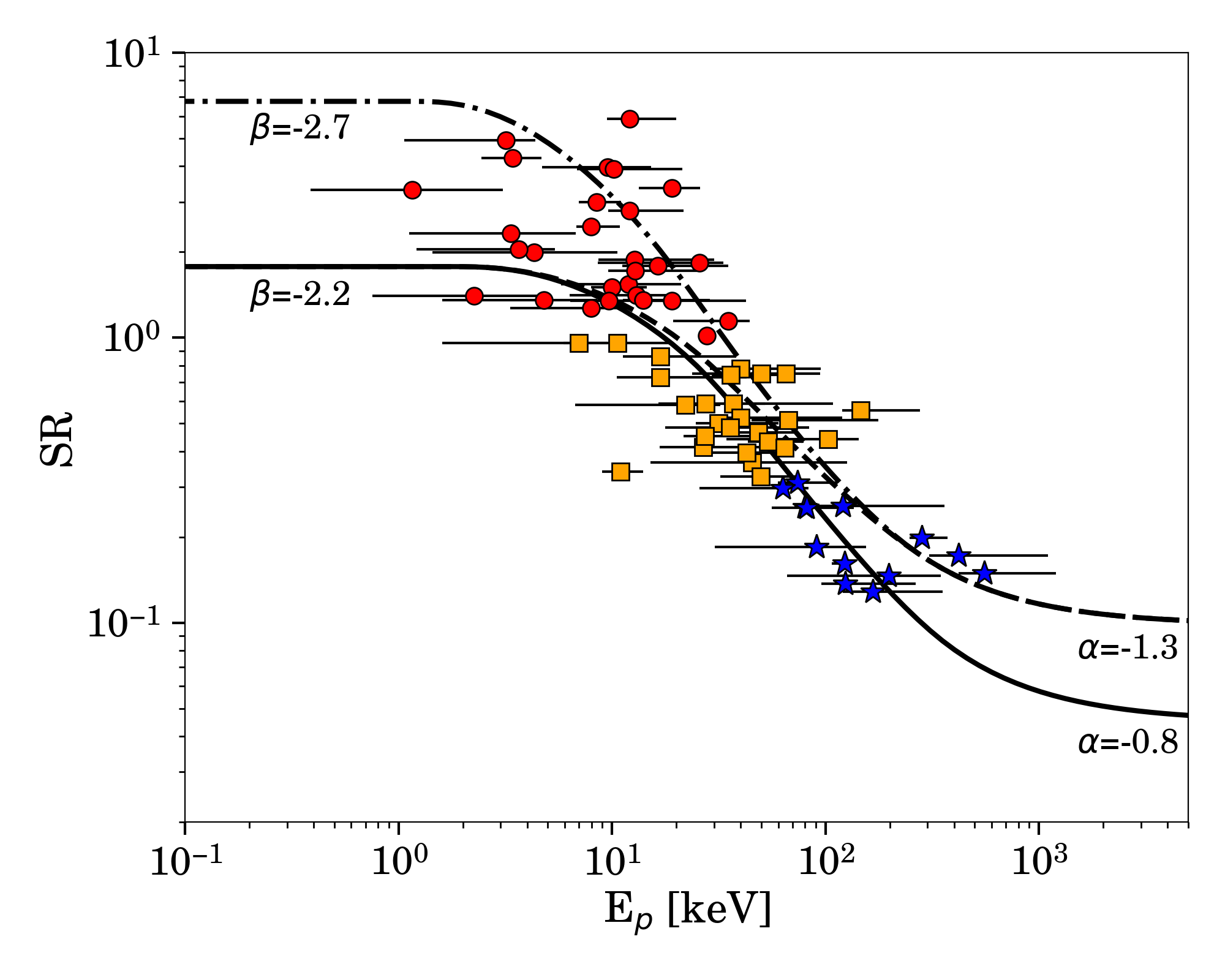}
\caption{Softness ratio as a function of the peak energy. GRBs ($\sr \le 0.32$), XRRs ($ 0.32 \le \sr \le 1$), and  XRFs ($ \sr \ge 1$) are represented in blue stars, orange squares and red circles respectively.
The dot-dashed line represents SRs for $\beta=-2.7$ and $\alpha=-1.3$. The dashed line represents SRs for $\beta=-2.2$ and $\alpha=-1.3$. Finally, the solid line represents SRs for $\beta=-2.2$ and $\alpha=-0.8$.}

\label{fig:Epeak_SR}
\end{figure}

In the case of $\beta$ (see Fig. \ref{fig:beta}),  GRBs appear to show larger (flatter) values compared to softer events. The cases with $\beta > -2$ are physically meaningless (because of energy divergence), simply meaning that this parameter is poorly constrained in the high (flat) end.
In fact, most GRBs have an $\epc$ above 40 kev, where only two spectral data points are available to determine $\beta$. Indeed, the number of events with well constrained $\beta$ decreases rapidly with $\epc$ approaching the upper
energy threshold (4 GRB vs 14 XRR and 12 XRF). In this regime, GRBs with steeper values of
$\beta$ will give less counts in the GRBM, and therefore tend to have a unconstrained value
of  $\beta$ and are not included in the computation of the average. 
In summary, the different spectral shape exhibited by the three classes, from the soft spectrum XRF to the hard spectrum GRB is determined primarily by $\epc$. This is clear from Fig. \ref{fig:Epeak_SR}, where the correlation between SR and $\epc$ is represented. The three black lines represent the SR for two different values of $\beta=-2.2$ and -2.7, and $\alpha=-1.3$ and -0.8. While $\epc$ produces the overall shape of the correlation, $\beta$ only influences SR at low $\epc$, and $\alpha$ at high $\epc$. 

\begin{figure}
 \includegraphics[width=0.5\textwidth]{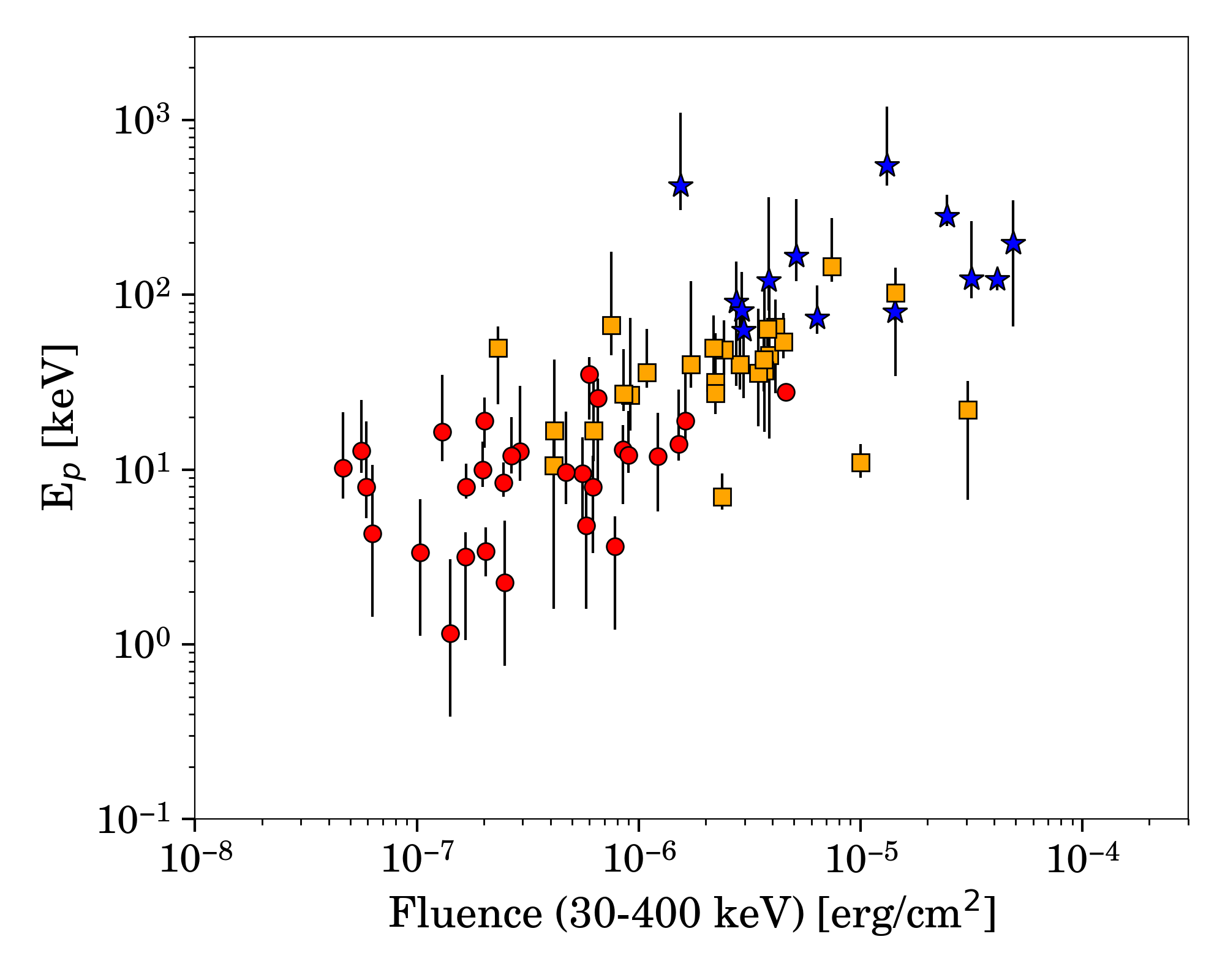}
\caption{Spectral peak energy as a function of the fluence in the 30-400 keV band. In red XRFs, in orange XRRs and in blue GRBs. }
\label{fig:ep_vs_fl}
\end{figure}

The observed 30-400 keV fluence shows a strong correlation with $\epc$, with XRFs being less energetic than GRBs (see Fig. \ref{fig:ep_vs_fl}). This is simply a consequence of the Band function: given that $\alpha$ and $\beta$ are on average independent of the class (and therefore of $\epc$), a large $\epc$ causes a large fluence in the band 30-400 keV.

\subsection{Duration}

\begin{figure*}
\centering
 \includegraphics[width=0.4\textwidth]{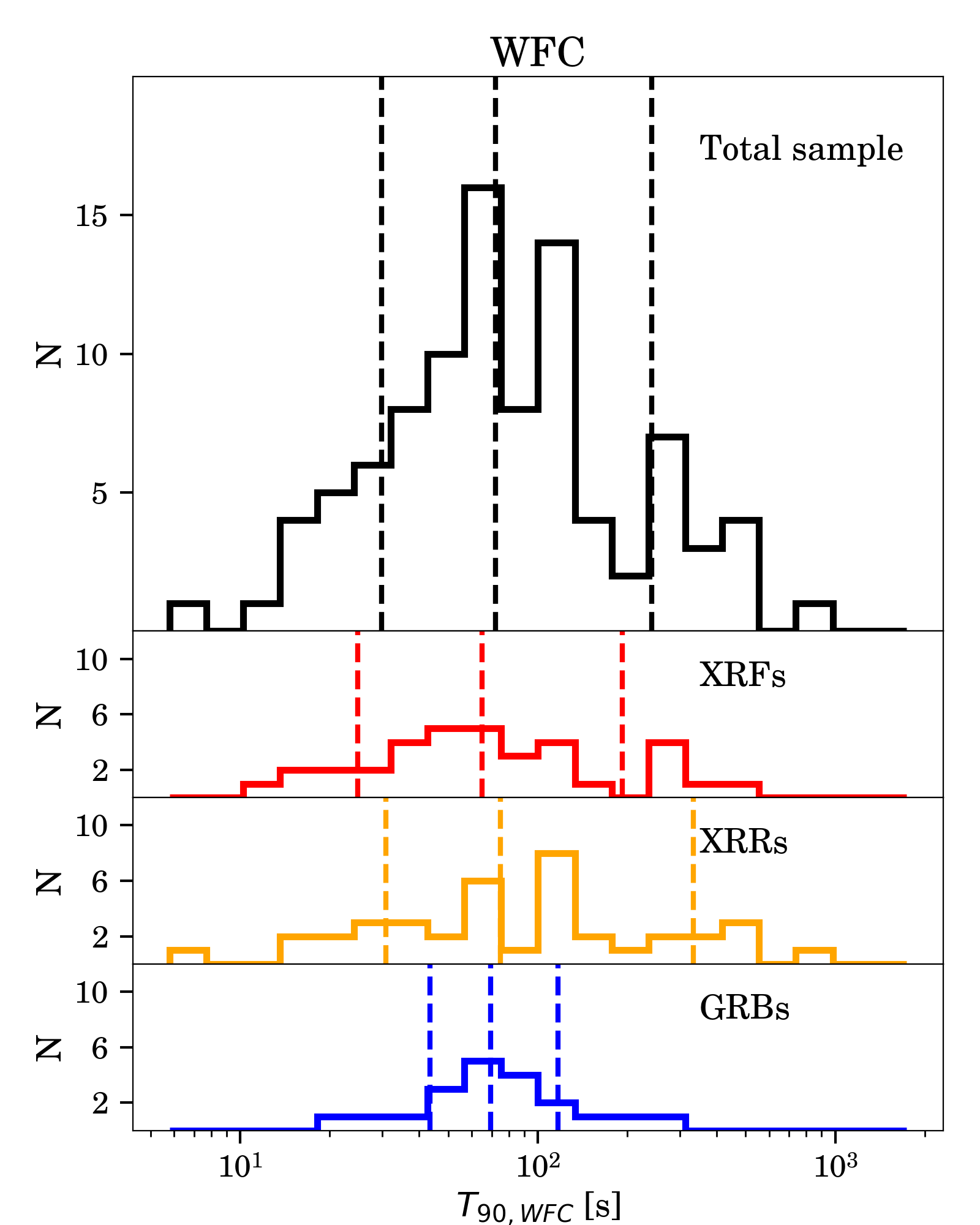}
\includegraphics[width=0.4\textwidth]{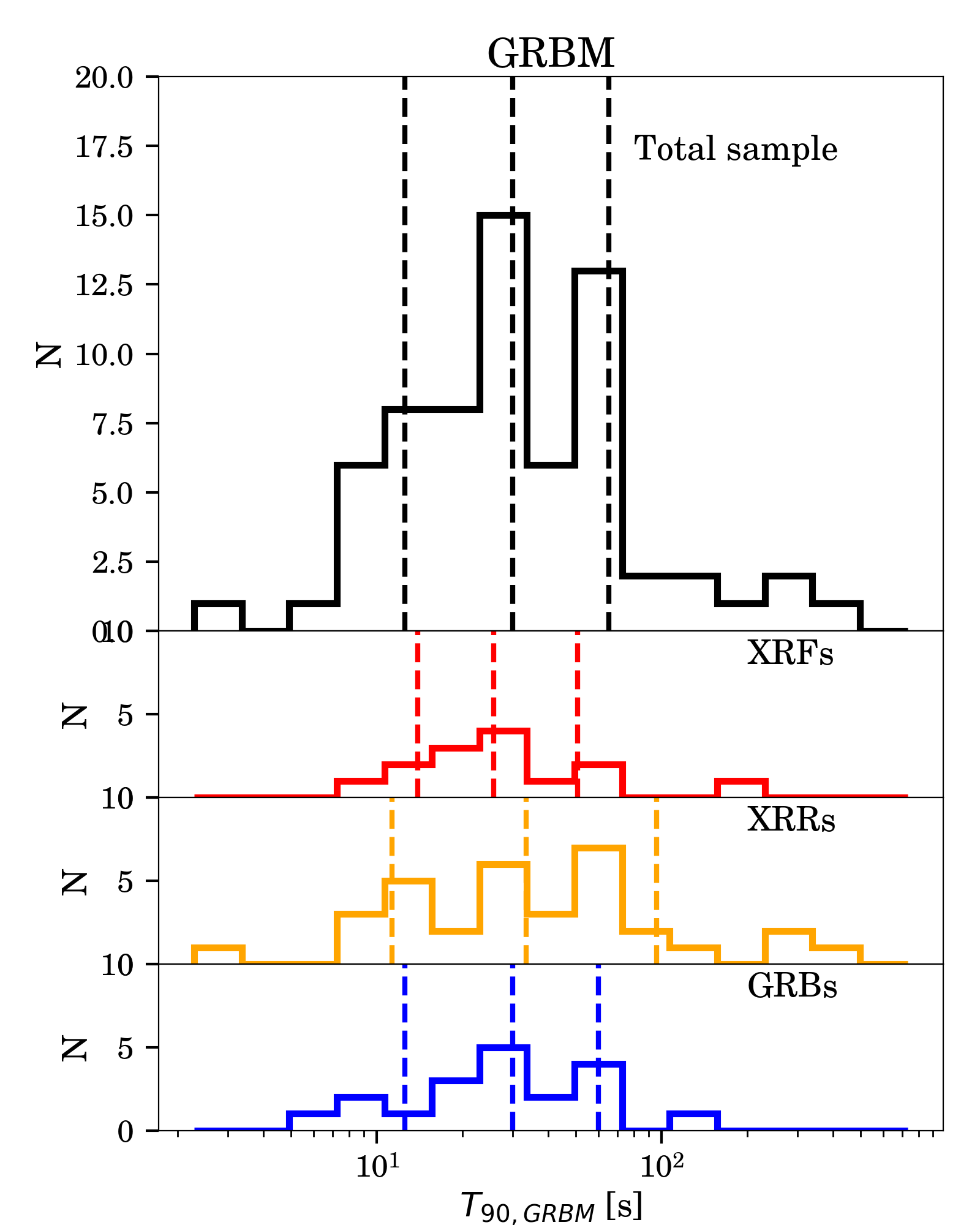}
  \caption{Left panel: WFC $\rm T_{90}$ for the full sample (black), for GRBs (blue), XRRs (prange) and XRFs (red). Vertical dashed lines represent median, 16th, and 84th percentiles. Right panel: same as left panel, for GRBM detected bursts.}
\label{fig:T90_WFC}
\end{figure*}

\begin{figure}
\center
 \includegraphics[width=0.5\textwidth]{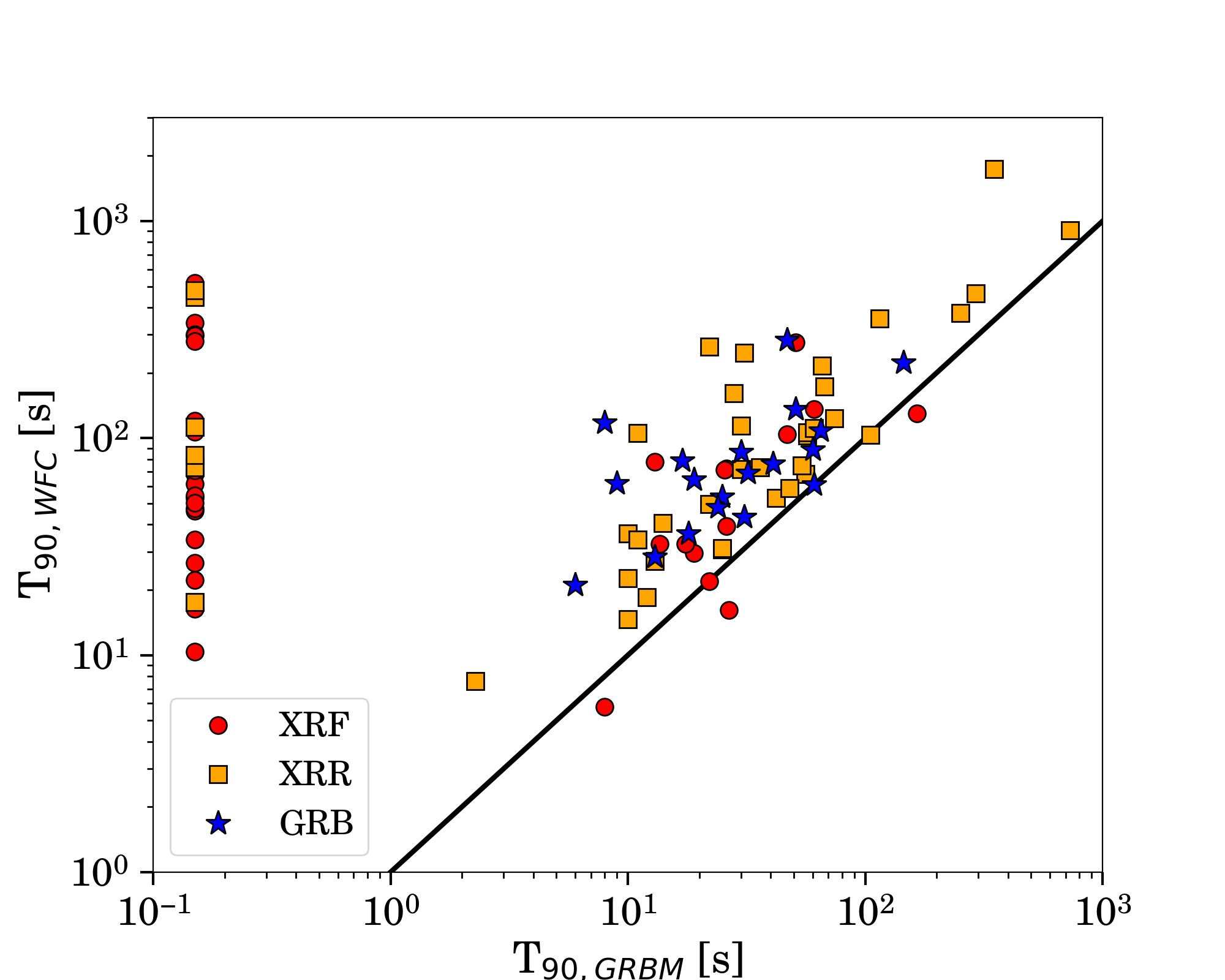}
\caption{Distribution of the events as function of $T_{90}$ in the GRBM and WFC. GRBs, XRRs and XRFs are identified with star, square, and filled circle symbols. The group of points at T$_{90}$(GRBM)=0.15s are those with no detection in the GRBM. The black solid line represents $T_{90, WFC} = T_{90, GRBM}$.}
 \label{fig:duration_GRBM_WFC}
 \end{figure}

We did not find any significant difference in the time duration ($T_{90}$) of the three classes of
events in either the WFC (see Fig. \ref{fig:T90_WFC}) and the GRBM light curves. We note that there are 14 XRF and 34 XRR that are bright enough to be detected in the GRBM.  The median duration in the WFC for GRB,
XRR and XRF are respectively $70^{+50}_{-24}$ s, $75^{+230}_{-42}$ s, $72^{+130}_{-43}$ s (medians and 16th, 84th percentiles) and in the GRBM are $30^{+30}_{-18}$ s, $30^{+36}_{-19}$ s, $22^{+9}_{-6}$ s (medians and 16th, 84th percentiles). The duration of the event in the WFC bandpass is
significantly longer than in the GRBM. This behaviour is consistent with previous
observations, showing that the duration scales with $\sim E^{-[0.4-0.5]}$ \citep{Piro1998, Fenimore1995}.
In Fig. \ref{fig:duration_GRBM_WFC} we show the distribution of event durations in the WFC and in the GRBM of the three classes. 
The solid line corresponds to $T_{90, WFC}=T_{90, GRBM}$. 

\subsection{Precursors}
\label{sec:precursors}

A distinctive feature of precursors previously found in \sax GRBs is their softer spectrum as opposed to the main event (e.g. GRB 980519, GRB 981226, XRF011030, and GRB 011121, \citealt{Zand1999, Frontera_2000, Galli2006, Piro_2005}). We thus carried out a  search for events that both precedes in time and whose spectrum is substantially softer than the main train of pulses. 
We first applied a time-driven selection by searching for the presence of emission in the WFC before the GRBM onset. The onset of the WFC emission is obtained by the extensive search in the WFC full data set derived in paper I. We required a  time difference  larger than 10 s, safely above the error in the determination of the onset of the pulses in the two instruments. We have further checked the light curves to confirm the presence of the precursor. We then verified that   $\srobs$ during the precursor is larger than the one of the main pulse. 
We note that this method is conservative because, while providing a robust sample selection, it excludes limiting cases of precursors that are either strong enough to be detected in the GRBM, or to be recognized in the WFC by their soft spectrum (as an example XRF 011030) or that are at the limit  of  the threshold for the blind search throughout the full WFC data  (GRB 981226, \citealt{Frontera_2000}). In these two cases the precursors started respectively about 300s, and 80s before the main peak. 

\begin{figure*}
\centering
 \includegraphics[width=0.23\textwidth]{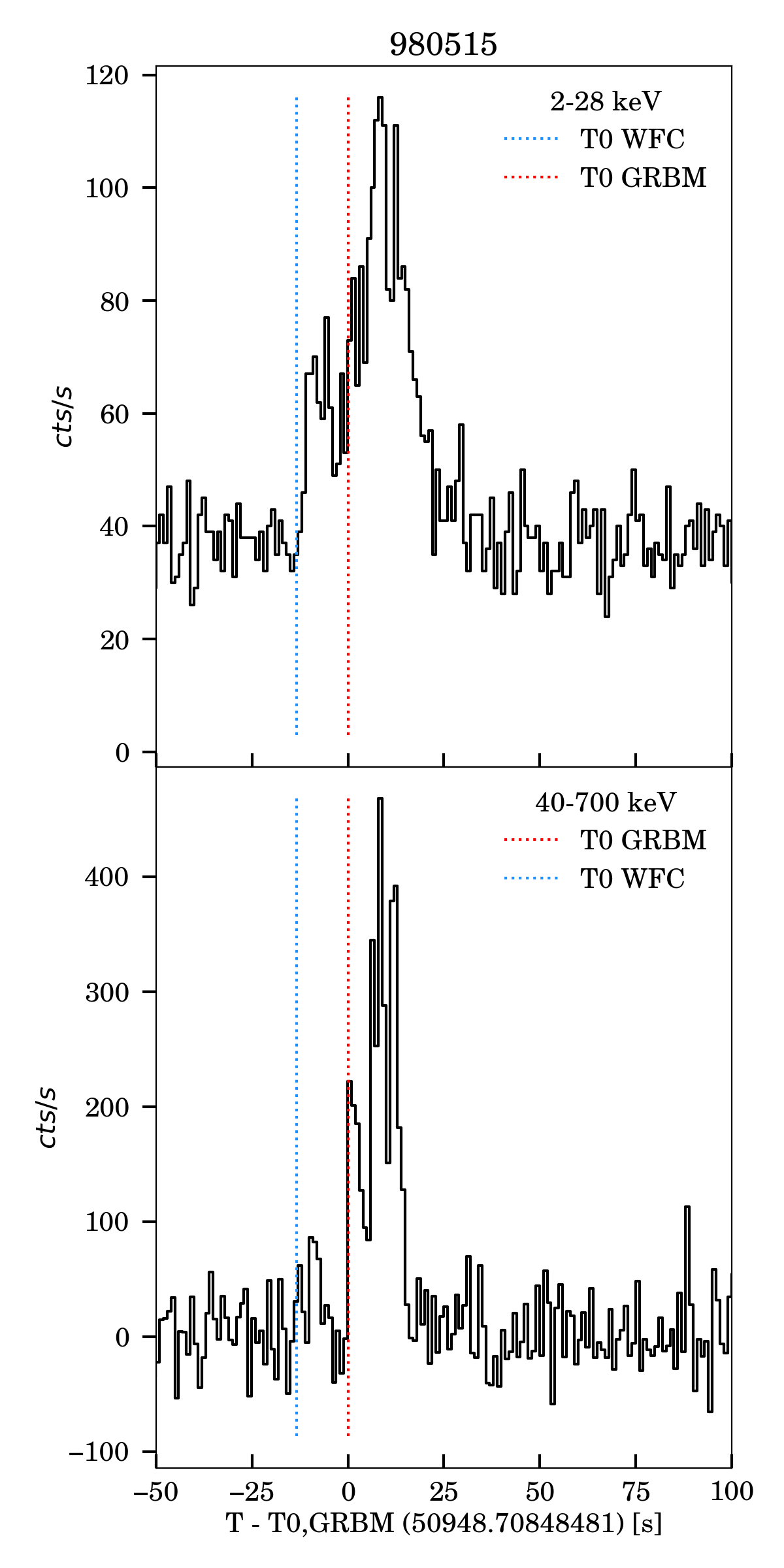}
 \includegraphics[width=0.23\textwidth]{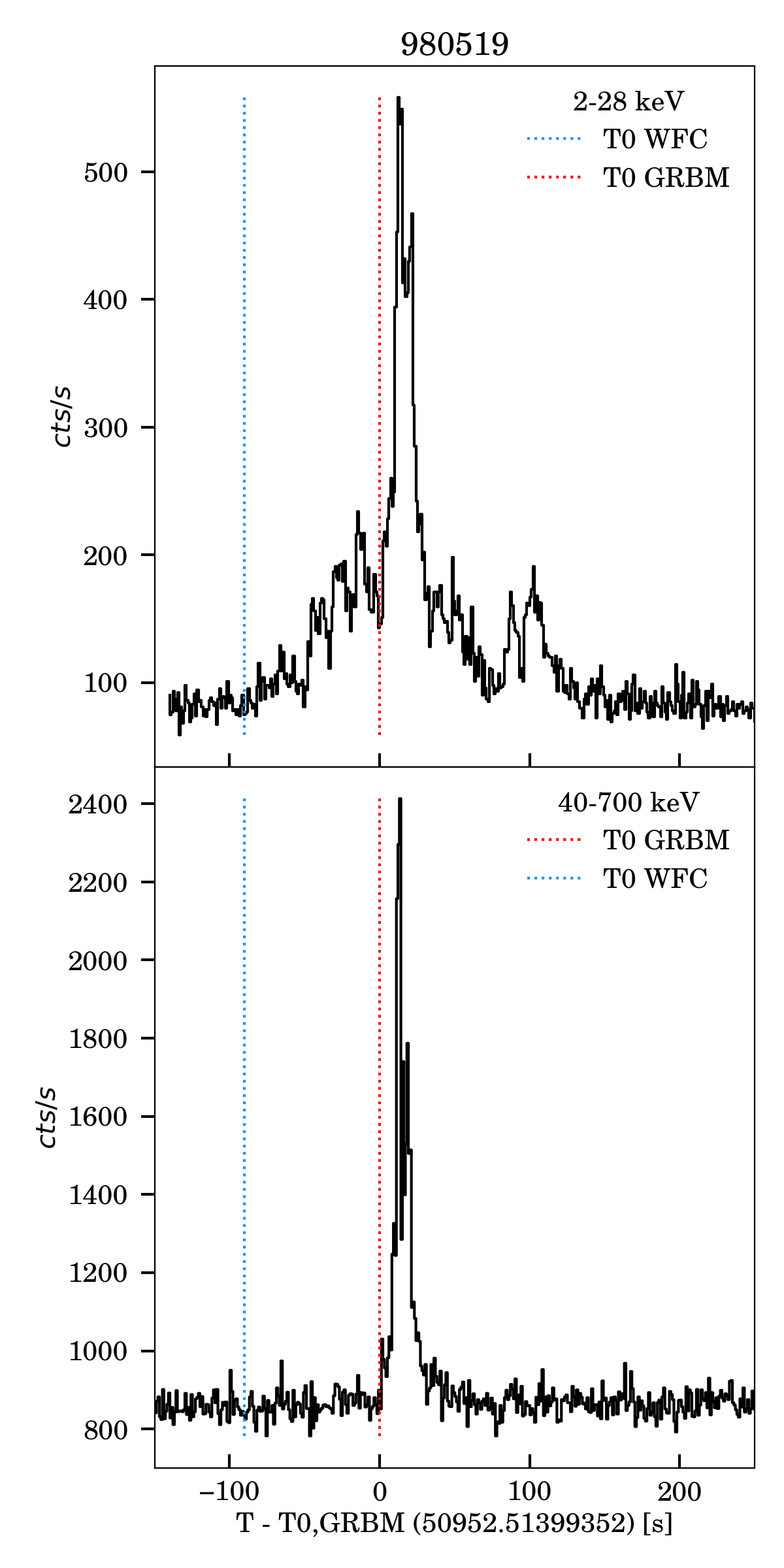}
 \includegraphics[width=0.23\textwidth]{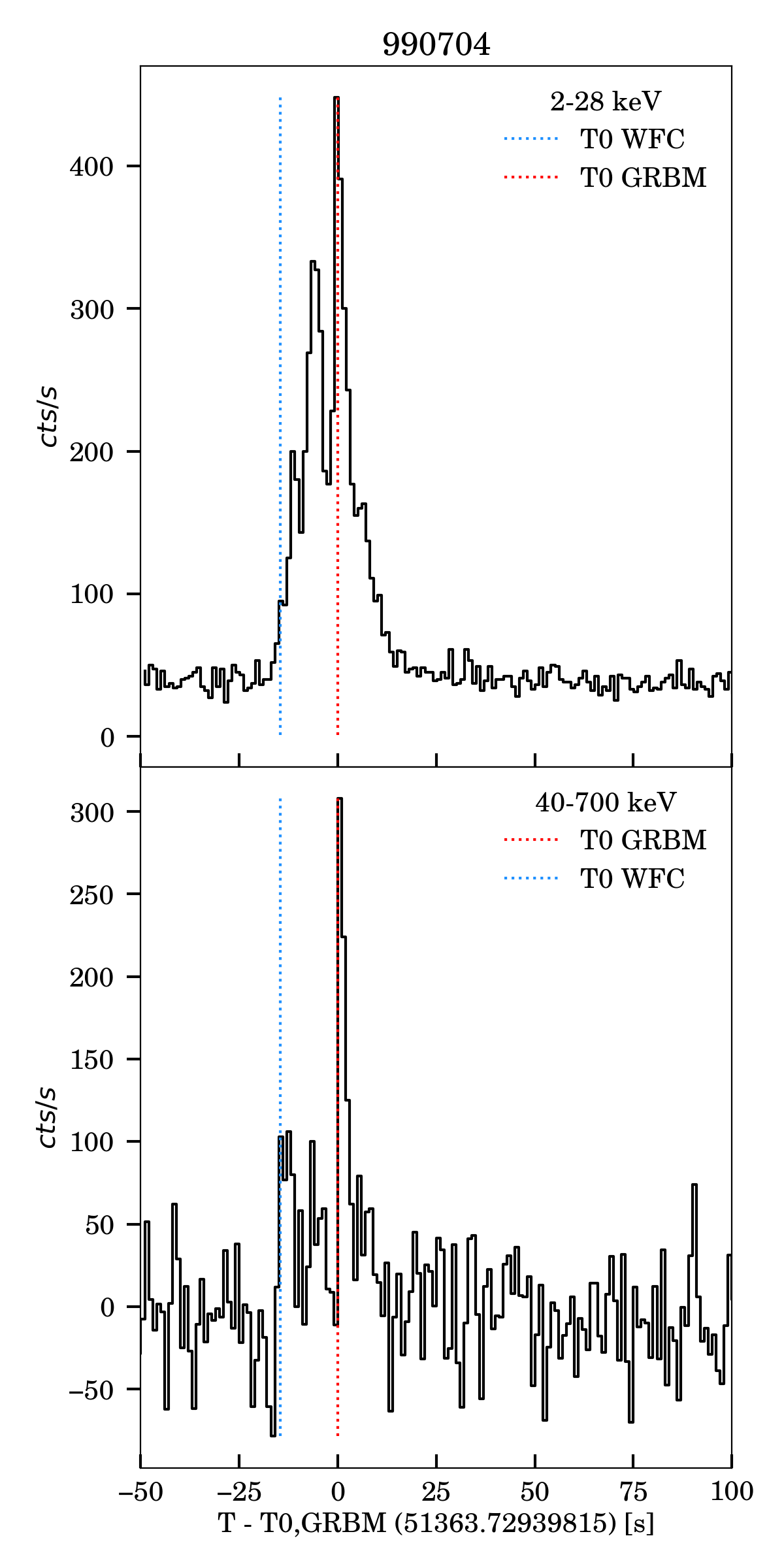}
 \includegraphics[width=0.23\textwidth]{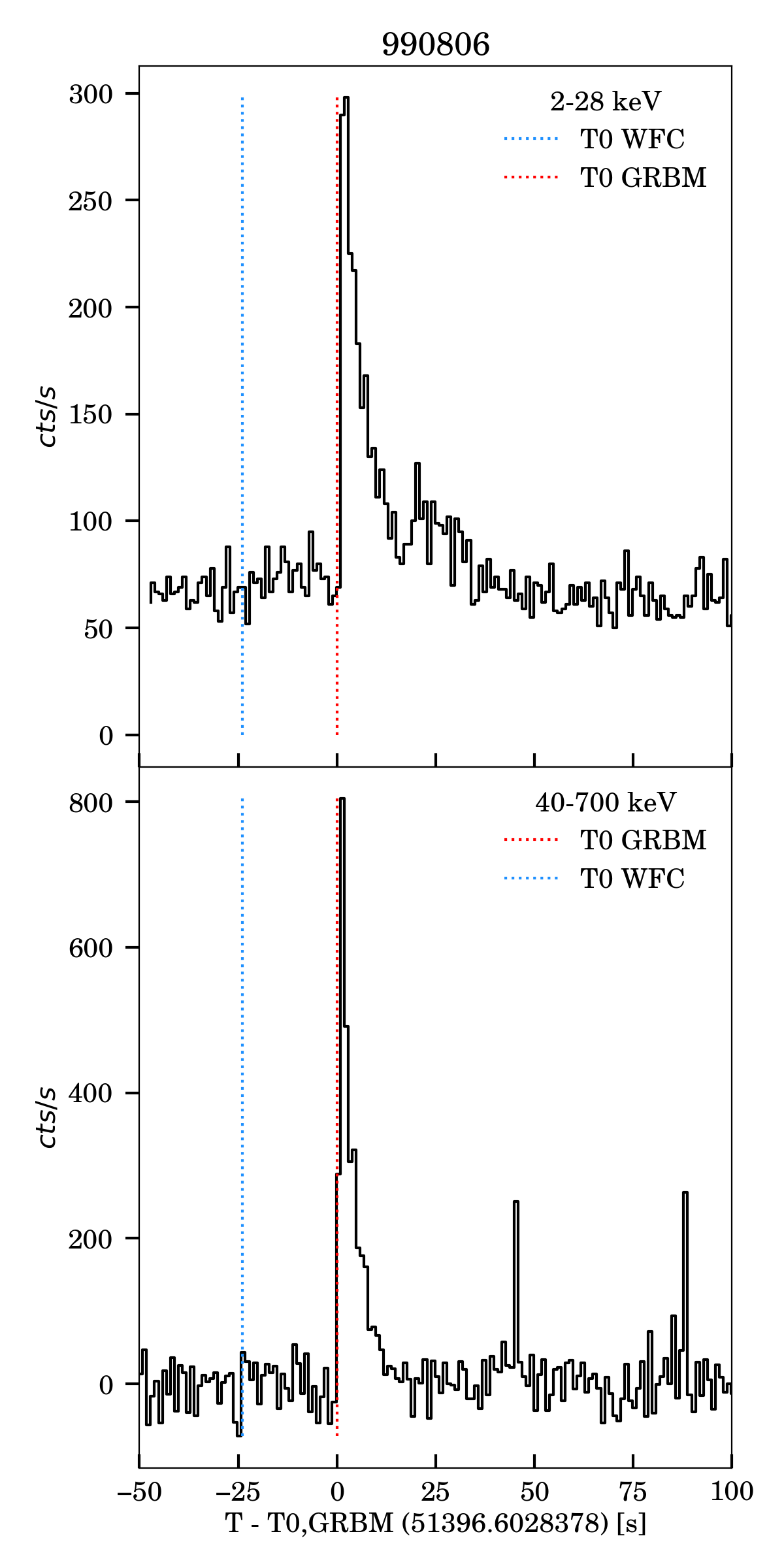}
 \includegraphics[width=0.23\textwidth]{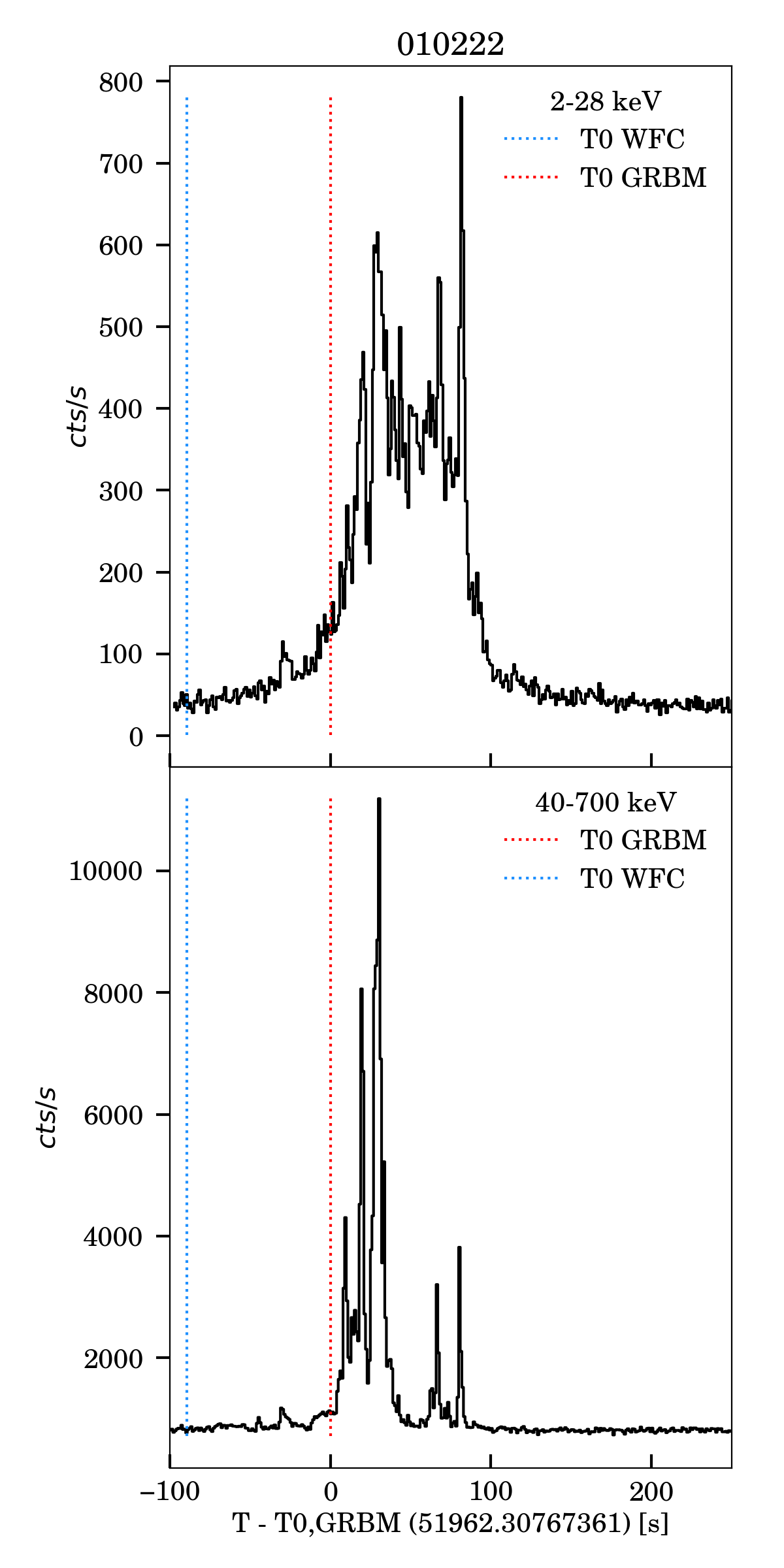}
\includegraphics[width=0.23\textwidth]{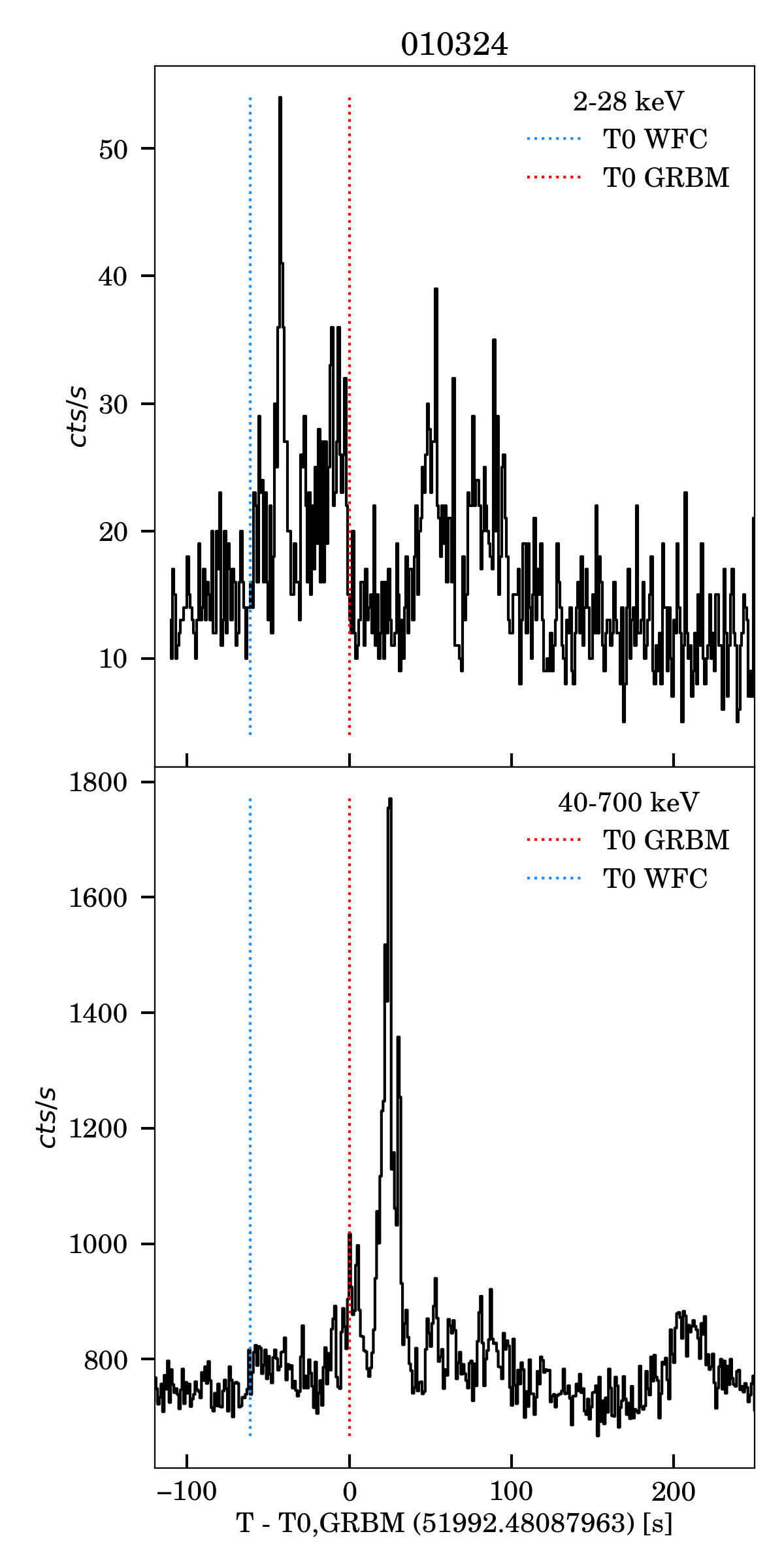}
 \includegraphics[width=0.23\textwidth]{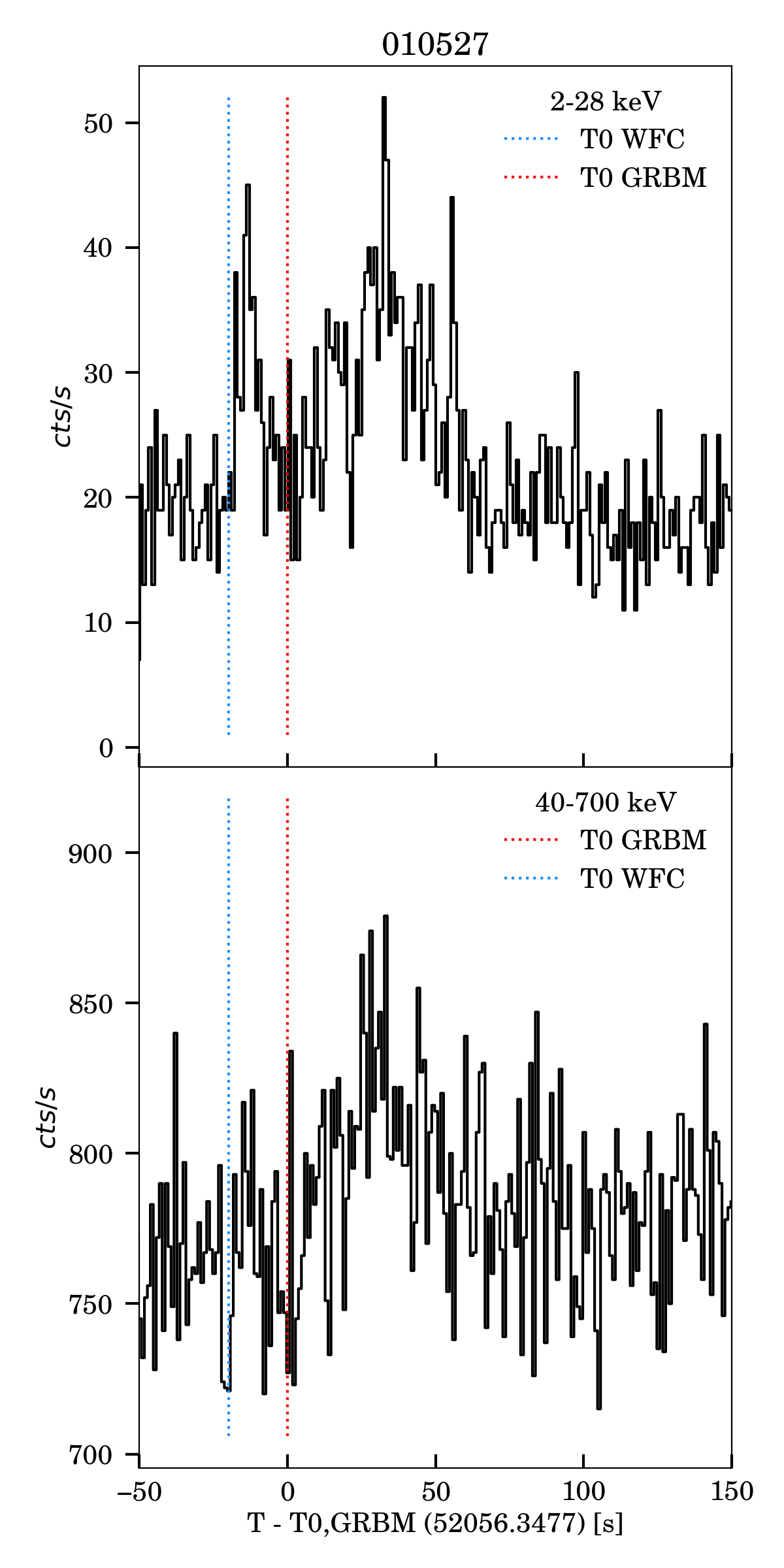}  
 \includegraphics[width=0.23\textwidth]{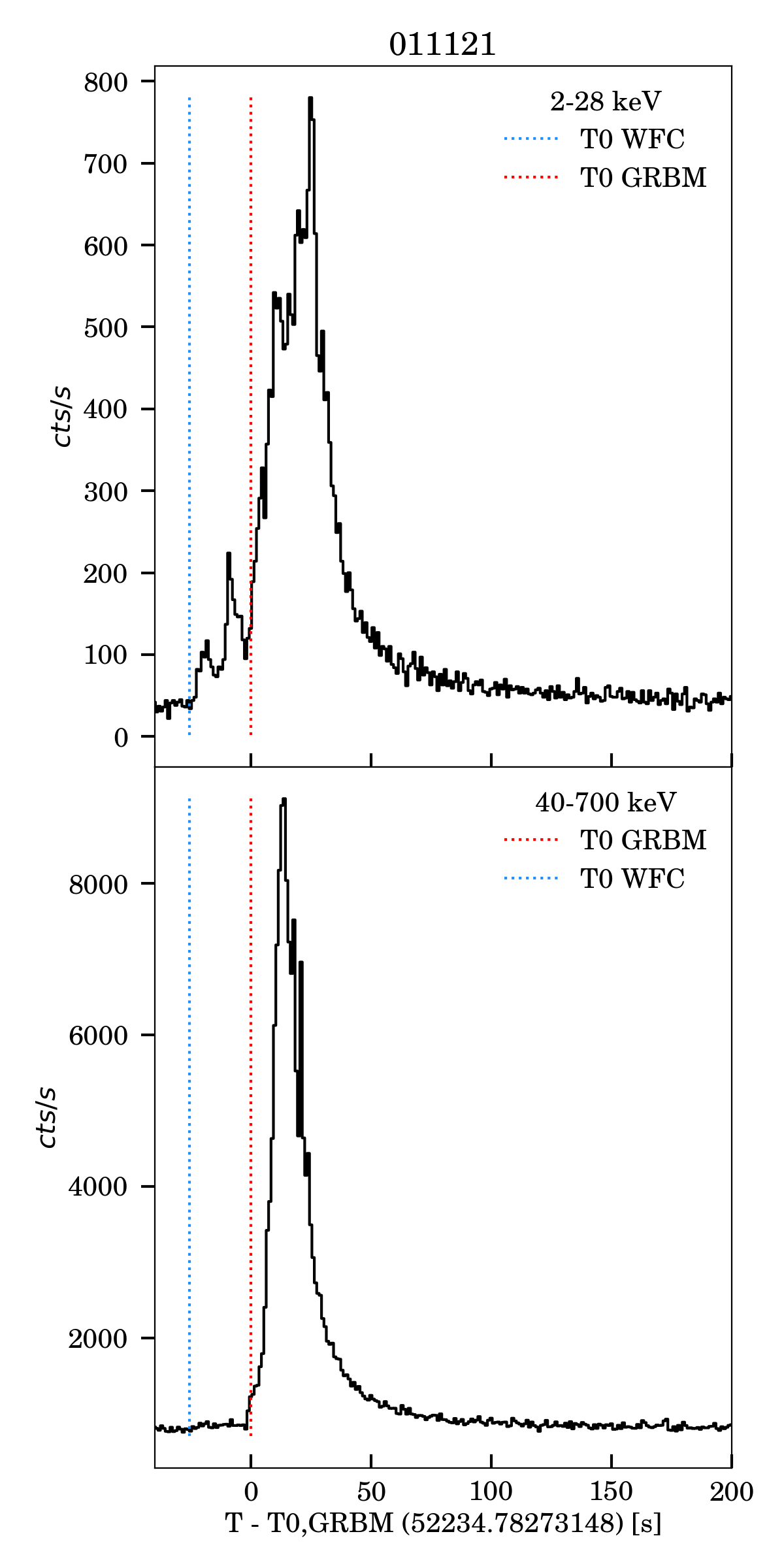}
 \includegraphics[width=0.23\textwidth]{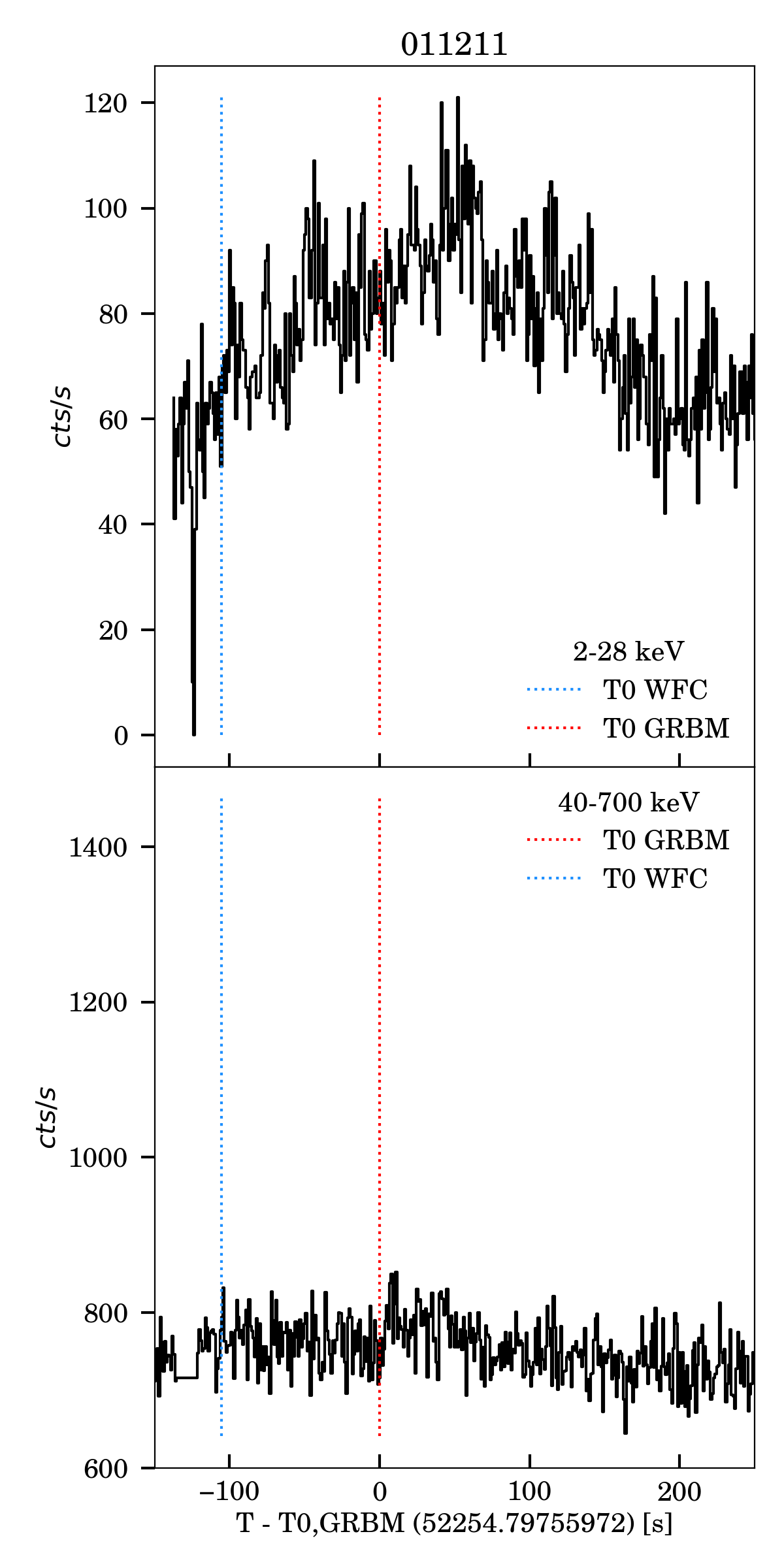}
  \caption{Light curves for the nine precursors found in \sax bursts. $\rm T_0$ is the onset time of the GRBM signal (see \citealt{Frontera2009}). GRB 010222 and 010324 there seem to be a marginal signal also in the GRBM during the WFC precursor. In these cases, the $\srobs$ of the precursor (see Table \ref{table:precursors}) is still softer than the main emission peak, which follows our definition of precursor.}
\label{fig:precursor_LCs}
\end{figure*}

Of the 67 bursts with a GRBM detection, we find  9 bursts ($\approx14\%$) with a precursor emission (see Table \ref{table:precursors}). Out of these, 2 were previously recognized (GRB 980519 and GRB 011121, \citealt{Zand1999, Piro_2005}).
We note that for XRR 010222 and 010324 there is close to marginal signal also in the GRBM at the time of the WFC precursor (see Fig. \ref{fig:precursor_LCs}). In these cases, the $\rm \srobs$ of the precursor is still softer than the main emission peak (see Table \ref{table:precursors}). 

\begin{figure*}
 \includegraphics[width=\textwidth]{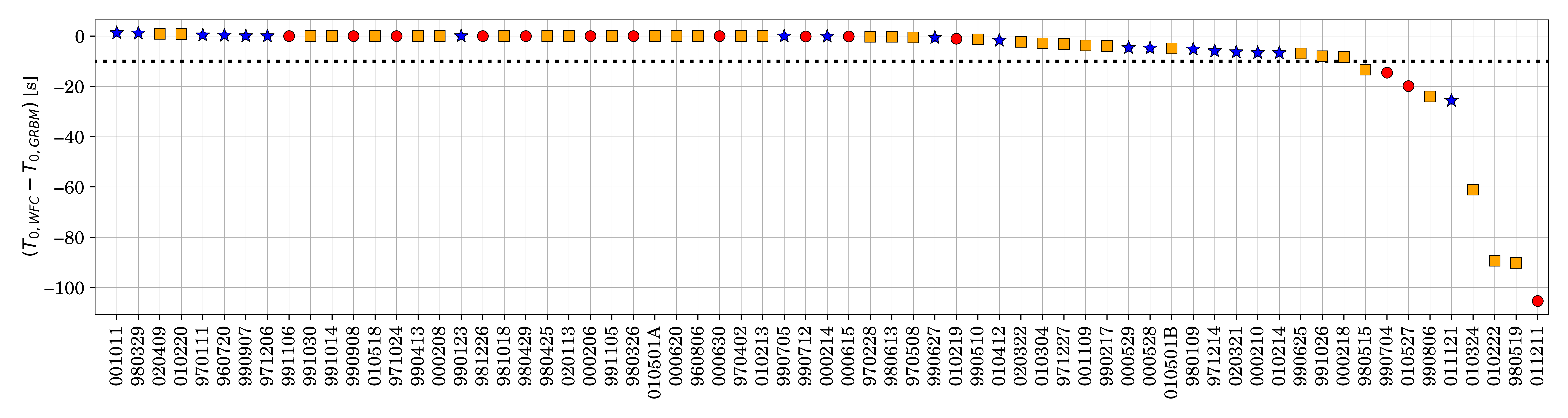}
\caption{$\Delta T = T_{0, WFC} - T_{0, GRBM}$ for each \sax burst. Bursts with identified precursor emission have $\Delta T > 10$ s (dotted line). GRBs, XRRs and XRFs are represented with blue stars, orange squares and red circles respectively.}
 \label{fig:precursors_deltaT}
\end{figure*}

In our sample  precursors precede the GRB main pulse  from  14 s to 105 s (see Fig. \ref{fig:precursors_deltaT} and Table \ref{table:precursors}).  Interestingly, there is no significant difference in the fraction of precursors per class, with 3 out of 14 XRFs, 5 out of 34 XRRs and 1 out of 19 GRBs (see also Table \ref{table:numbers}), all about 10-15$\%$.
On the contrary we do not find presence of post-cursor X-ray emission, i.e. events where the onset of the WFC first pulse follows the gamma-ray event by more than about 10 s.

\begin{table}
\caption{Precursors for \sax bursts. } 
\label{table:precursors}     
\centering                         
\begin{tabular}{c c c}      
\hline\hline               
GRB & $\Delta T$ [s] & $\rm SR_{obs, prec} [10^{-2}]$ \\   
\hline                   
   XRR 980515 & 14 & 76 \\
   XRR 980519 & 90 & 1900 \\
   XRF 990704 & 15 & 361 \\
   XRR 990806 & 24 & 500 \\
   XRR 010222 & 89 & 13 \\
   XRR 010324 & 61 & 23 \\
   XRF 010527 & 18 & 49 \\
   GRB 011121 & 26 & 1204 \\
   XRF 011211 & 105 & 9 \\
\hline                                   
\end{tabular}
\tablefoot{
    The GRB name is reported in the first column, while the onset time before the main peak and the observed softness ratio of the precursor are written in the second and third columns.}
\end{table}

\section{Properties of the associated afterglows}
\label{sec:properties_afterglow}

\begin{figure}
\centering
 \includegraphics[width=0.5\textwidth]{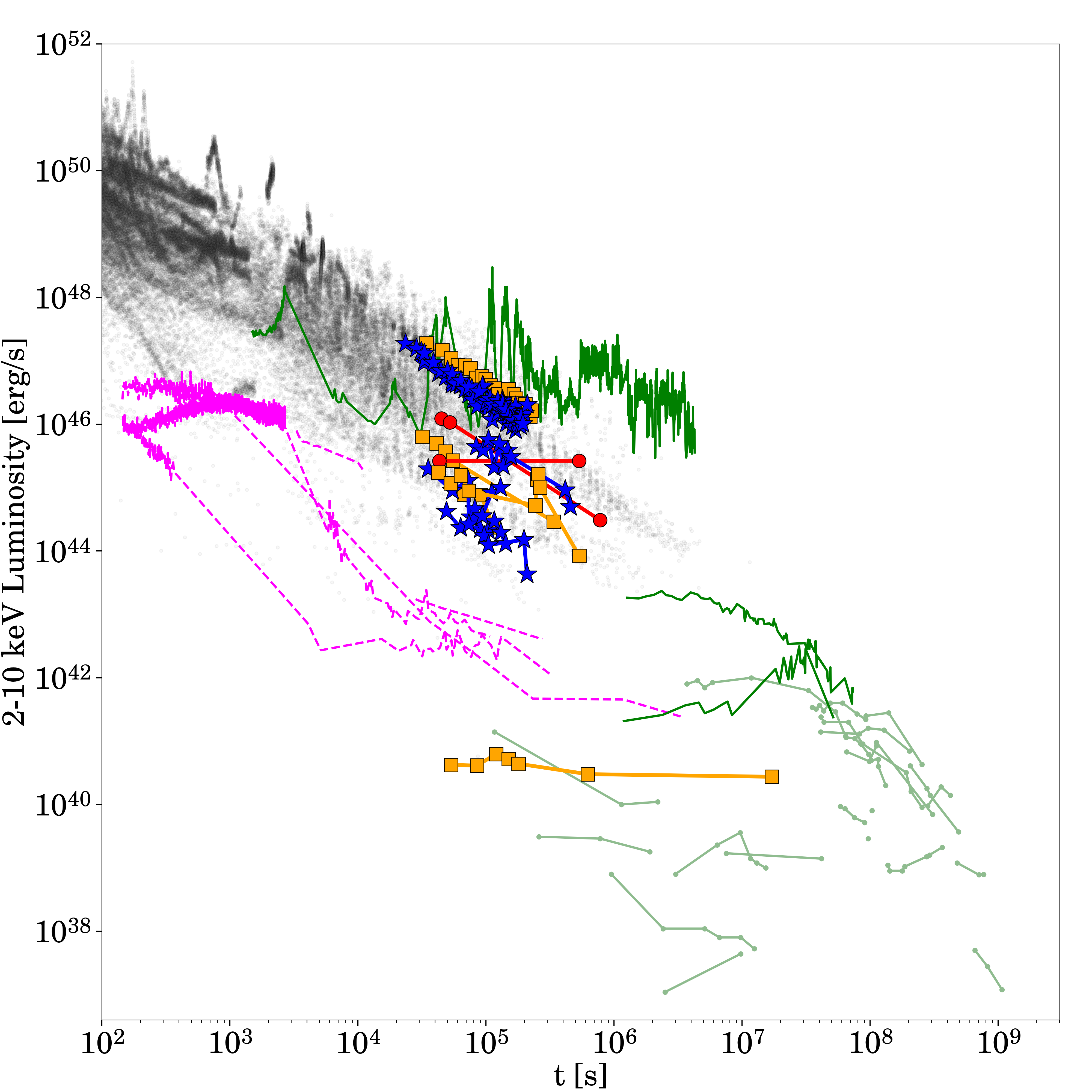}
  \caption{Afterglow luminosity of \sax bursts as a function of rest-frame time.  XRF, XRR and GRBs are represented in red circles, orange squares and blue stars}. In grey the \swift sample of long GRBs, in green examples of Tidal Disruption Events \citep{Burrows2011, Gezari2017, Mummery2020}, in light green SN \citep{Ross2017}, and in magenta low-luminosity GRBs. 
\label{fig:afterglow_LCs}
\end{figure}

\sax sent a real time alert to the community for 53 events, over the total of 96 in our sample. Of this subsample, 37 (70$\%$) were followed up in the X-rays and afterglow was found in 90$\%$ of the cases (32 events). Of the 52 events followed up by optical telescopes, 35$\%$ (18 events) had an afterglow. Finally, of the 36 events followed up in the radio, 33$\%$ (12) had a counterpart. Looking at the subclasses of XRFs, 78$\%$ had an X-ray afterglow, 29$\%$ had an optical afterglow and 29$\%$ had a radio afterglow. For the XRR population, the afterglow was discovered in the X-rays in the 93$\%$ of the cases, in the optical 45$\%$, and in the radio 42$\%$. Finally, for GRBs, 84$\%$ had an X-ray afterglow, 33$\%$ had an optical afterglow, and 33$\%$ had a radio afterglow. These numbers are also reported in Table \ref{table:numbers}.
In Fig. \ref{fig:afterglow_LCs} we show the 2-10 keV afterglow light curve luminosity for the \sax bursts with a redshift identification. Our bursts have luminosities consistent with the GRB population  (the \swift GRB population is represented in shaded grey in the Figure), except for GRB 980425. The latter has an X-ray luminosity 2-3 order of magnitudes lower than canonical GRBs, and about 1 order of magnitude with respect to Low-Luminosity GRBs in magenta. Instead, it has similar luminosity to SNe, represented in light green.

\section{Discussion}
\label{sec:discussion}

\subsection{Progenitors of XRF vs GRBs}

We find that XRFs, XRRs and GRBs have similar durations in the prompt emission, both in X-rays and gamma-rays. They also show spectral properties consistent with a single population of events with different $\rm E_{p}$, and, for the most part, follow the $\rm E_{p,i}-E_{iso}$ (with $E_{p,i}$ the peak energy corrected for redshift and $E_{iso}$ the isotropic-equivalent energy, \citealt{Amati2006}) relation, as we will show later. This is in agreement with results from \hete, \swift and Fermi bursts \citep{Sakamoto2005, Sakamoto2008, Bi2018, Katsukura2020}.
The broad-band afterglow detection rates are consistent within the three classes \citep[see also][]{DAlessio2006}. There seems to be some slight difference in the afterglow temporal behavior, with XRFs having shallower decays with respect to GRBs \citep{Sakamoto2008, Katsukura2020}, and lower luminosities, even if still consistent with the GRB population.
In addition to our findings we also note that they share similar redshift distributions \citep{Bi2018}, and XRF host galaxies are similar to long GRB \citep{Bloom2003, Chen2021}.

\begin{figure}
\centering
 \includegraphics[width=0.5\textwidth]{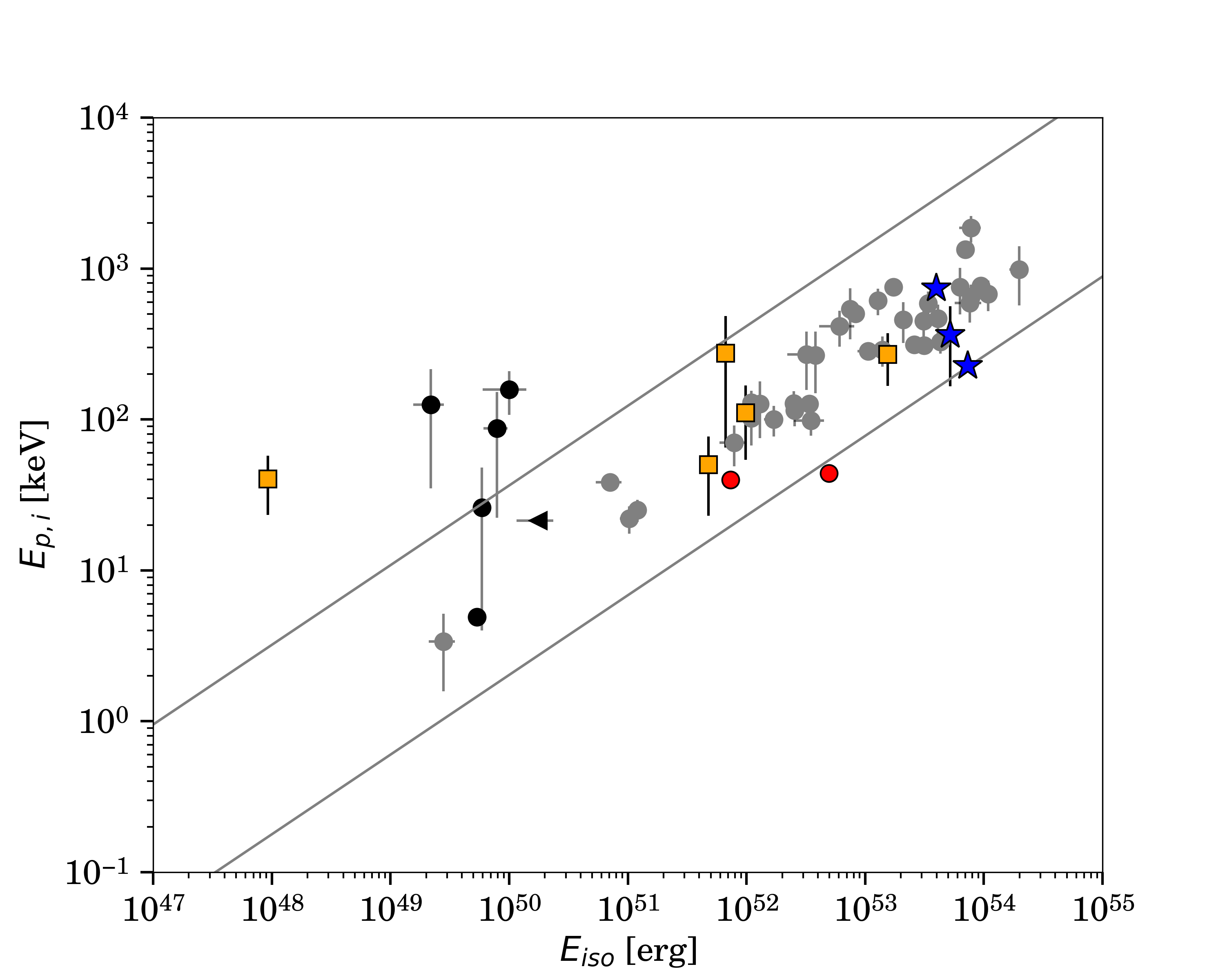}
  \caption{In gray $\rm E_{p,i}$ (the peak energy of the GRB's prompt emission spectrum in the burst's rest frame) and $\rm E_{iso}$ (isotropic-equivalent energy) relation from \citealt{Amati2006} for long GRBs, lines represent the 3$\sigma$ region. In color are represented our results for the \sax bursts.  Only the bursts for which $\rm E_{iso}$ could be constrained are included in the Figure.  Some of our \sax bursts are also present in \citealt{Amati2006}, in these cases they are plotted only once, and are in agreement within errors. In black examples of low luminosity GRBs \citep{Pian2006, LLGRB031203, LLGRB100316D, LLGRB120422, LLGRB201015A}.}
\label{fig:Epeak_Eiso}
\end{figure}

Although the exact origin of XRFs remains under debate,  their numerous commonalities with GRBs found suggest that they may share a common progenitor.
In fact, many XRFs are associated with SNe, the smoking gun of the collapsar scenario. XRFs seem to have a higher rate of association with SNe compared to GRBs \citep{Bi2018}. However, there could be a bias due to their low luminosities: XRFs tend to be observed at closer distances, which facilitates the search for a SN. Several models, in the collapsar context, have been proposed to explain the observed prompt properties. They could be high redshift GRBs \citep{heise}, but this can be excluded as the only source of XRF, as the redshift distribution found for \swift XRFs has a mean redshift of $\approx1$ \citep{Bi2018}. 
Another possible scenario is off-axis GRBs (uniform jet in \citealt{yamaz02,yamaz03,yamazaki04}, Universal Power-law jet in \citealt{lamb05}, Gaussian jet in \citealt{zhang03}, ring shaped jet in \citealt{eichler}), the observer viewing angle is outside the jet's cone, so the gamma-rays emitted by the jet are observed in the soft X-ray because of the Doppler beaming.  
Dirty fireballs, also called choked jet scenario \citep{dermer,barr}, are also expected to produce a softer prompt emission with respect to canonical GRBs. In this case, the ejecta that pierces through the stellar envelope is not an ultra-relativistic jet, but an ejecta with small Lorentz factor ($\Gamma << 300$), either jetted or spherical (also called cocoon). 
Another popular interpretation is the shock breakout radiation \citep{Campana2006, Soderberg2008}, which might also be spherically symmetric and is simply due to the break out of a relavistic shock, whose radiation peaks in the soft X-rays \citep{Nakar_Sari2012}. It should be noted that such radiation should be purely thermal, which is not commonly observed in GRBs.
Other possibilities include multi-components systems, including collimated and spherical ejecta, whose observed emission changes depending on the inclination of the system \citep{huang04}.

The fact that XRFs could be produced by a number of different scenarios could be confirmed by the scatter in the $\rm E_{p,i}-E_{iso}$ relation \citep{Amati2002, Amati2006}, larger for soft and low energy bursts (both XRR and XRF regions), see Fig. \ref{fig:Epeak_Eiso}. In this plane, XRFs and XRRs overlap with the class of Low-Luminosity GRBs \citep[LL-GRBs, ][]{Pian2006, LLGRB031203, LLGRB100316D, LLGRB120422, LLGRB201015A}. While there are some bursts like XRF 060218, a low-luminosity, soft, long lasting events, which are fully compliant with the $\rm E_{p,i}-E_{iso}$ relation, there are also outliers, the best example being the \sax GRB 980425, an XRR. Deviations from the Amati relation are explained by dirty fireballs and off-axis GRBs \citep{Xu2023}, which cannot follow the same relation of canonical on-axis GRBs.

\begin{figure}
\centering
 \includegraphics[width=0.5\textwidth]{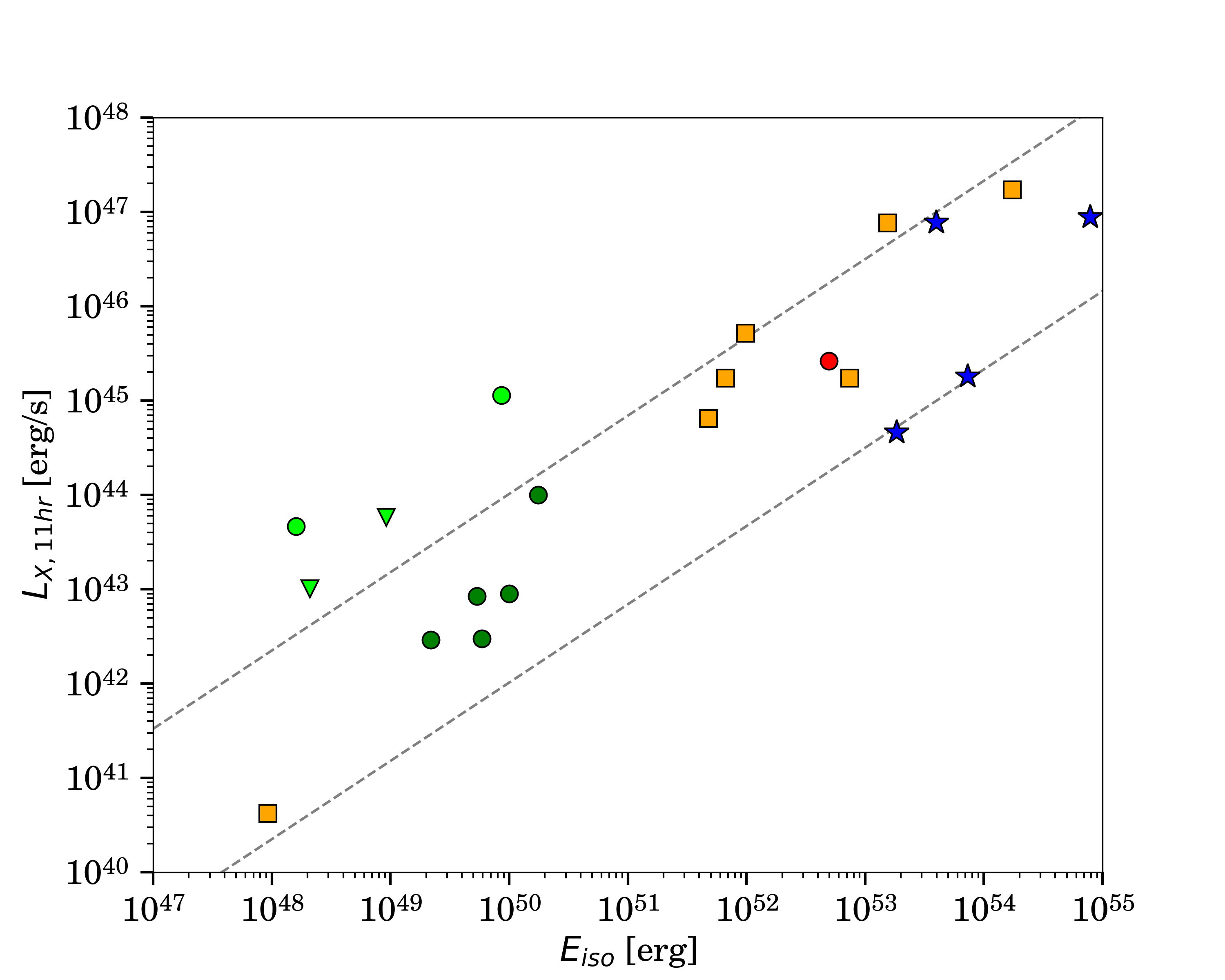}
  \caption{Rest-frame X-ray luminosity of the afterglow at 11 hrs versus $\rm E_{iso}$. In blue stars, orange squares and red circles the GRBs, XRRs and XRFs of the \sax sample. In green circles Low Luminosity GRBs. Some of the EP events with low $\rm E_{iso}$ are represented in lime (upper limits for EP250108a, \citealt{EP250108a_Li2025}, and EP250827b, \citealt{EP250827b_Srinivasaragavan2025}, and datapoints for EP240414a,  \citealt{Sun2025, Zheng2025}, and EP250304a, \citealt{cotter2026}). Grey dashed lines represents \swift GRB population \citep{DAvanzo2012}. }
\label{fig:Lx_Eiso}
\end{figure}

Another interesting relation is the X-ray afterglow luminosity at 11 hr ($\rm L_{X,11hr}$) and $\rm E_{iso}$, represented in Fig. \ref{fig:Lx_Eiso}. For \swift GRBs, \citet{DAvanzo2012} find a correlation, represented by dashed lines. We find that the \sax bursts (represented with colored squares) follow this relation. The position of GRB 980425 catches the eye, being at the lower-left end of the diagram. This somewhat confirms that 980425 is a LL-GRB, even though it has an $\rm L_{X, 11hr}$ and $\rm E_{iso}$ even lower than the other LL-GRBs (in green dots), as noted previously in this work. The gap between 980425 and the other LL-GRBs is recently starting to be populated thanks to EP. Indeed, EP-WXT has a good sensitivity in the soft X-ray range (0.5-4 keV), perfectly suited to discover soft and low-luminosity transients. EP250108a \citep{EP250108a_Li2025, EP250108a_Eyles-Ferris2025, EP250108a_Rastinejad2025, EP250108a_Srinivasaragavan2025} and EP250827b \citep{EP250827b_Srinivasaragavan2025} are represented with lime inverted triangles, as no afterglow was detected in the X-rays, while datapoints represent EP240414a  \citep{Sun2025, Zheng2025, Hamidani2025, vanDalen2025}, and EP250304a \citep{cotter2026}.

The $\rm L_{X, 11hr} - E_{iso}$ relation provides constraints on the off-axis model for XRF and LL-GRBs.
Depending on the off-axis angle, the afterglow light curve exhibits an initial rise, followed by peak and decay, when finally the jet becomes visible at high latitudes thanks to the declining Doppler beaming. After the peak, corresponding to the jet break, the afterglow emission recovers the on-axis decay behaviour.
Thus, in order to obey the observed relation for our sample, any jet break should take place before 11 h.
Indeed, \sax and \hete XRF and XRR afterglows never show early rising slopes \citep{DAlessio2006}, which would be a very strong hint for an off-axis jet. 
By requiring that the peak of XRF and XRR afterglow light curves arrives earlier than 11 hr ($\sim0.5$d), we can estimate an upper  limit on the sum of the opening angle of the off-axis jet and the viewing angle. From the peak time of an off-axis afterglow \citep{Ryan2020}, we derive
\begin{equation}
    \theta_{\rm obs} +1.24\theta_{\rm C} < 0.25 \left( \frac{E_{50}}{n_0} \right)^{1/8} \ \rm rad
\end{equation}
where $\theta_{\rm obs}$ and $\theta_{\rm C}$ are respectively the angle between the observer line of sight and the jet axis, and the core opening angle of the jet. $E_{50}$ is the isotropic-equivalent energy of the jet in units of $10^{50} \rm erg$ and $n_0$ is the ISM density in units of $cm^{-3}$. Assuming for XRFs $( E_{50} / n_0 ) \simeq 1$, $\theta_{\rm obs} +1.24\theta_{\rm C} \leq 14^{o}$. This, considering that the mean opening angle for long GRBs is about 10$^o$, points to mildly off-axis jets, consistently with what derived by  \citet{DAlessio2006}. 

Even for \swift bursts, the X-ray afterglow of XRFs tends to be shallower than that of GRBs, not showing the typical jet break \citep{Sakamoto2008}. 
This would suggest an early jet break from a close viewing angle, as mentioned above, or an on-axis wide component, like a dirty fireball. However, the possibility of multi-component (or structured) off-axis jet cannot be ruled out. In this case, if the observer is off-axis with respect to the jet, but aligned with some external less energetic wings, the latter could dominate the emission, and produce a shallow decreasing light curve \citep{Gianfagna2025, Zheng2026}.

\subsection{Origin of precursors}

A distinctive feature of precursors is their softer spectrum as opposed to the main event.
Our conservative selection  requires the X-ray event to precede the gamma-ray event and be below the GRBM detection threshold, giving 9 precursors in our subsample of 67 GRBs. In addition, there are 2 other \sax events that were identified on the basis of their softness ratio \citep{Frontera_2000, Piro_2005}. So, the presence of this phenomenon is not negligible, being observed in $\gtrsim 13 \%$.
Recent observations by EP, characterized by a Wide Field X-ray Telescope \citep{Yuan2025} with an X-ray bandpass even softer that the \sax WFC are also finding similar precursors, i.e. with a spectrum softer than the main event and comparable (or somewhat larger) $\Delta T$s. So far, there are two cases published in literature: EP240315a \citep{Liu2025}, and
EP240801a/XRF240801 \citep{Jiang2025}.
Before EP and \sax, precursor soft X-ray emission was found also in \ginga GRBs. Of
121 GRBs, there is the one strong case, GRB 900126 with a black body spectrum for the precursor \citep{Murakami1991}, and two possible cases \citep{Murakami1992}. 
Other precursors have also been found in WATCH data \citep{Sazonov1998} and P78-1 data \citep{Laros1984}.

The origin of this phenomenon is likely associated to the emergence of the jet at the surface of the stellar envelope, that produces a low-energy, thermal dominated emission \citep{Piro_2005, Lazzati2005}
The early X-ray emissions may be also caused by a prior shock breaking out from the star envelope \citep{Bromberg2012, Lazzati2005, Piro_2005}, or some leading weak jets in a series of the intermittent jets \citep{Wang2007, Lopez2016, Geng2016} from the unstable accretion of the central engine.

\subsection {Comparison with Einstein Probe FXTs}

\begin{figure}
\centering
 \includegraphics[width=0.5\textwidth]{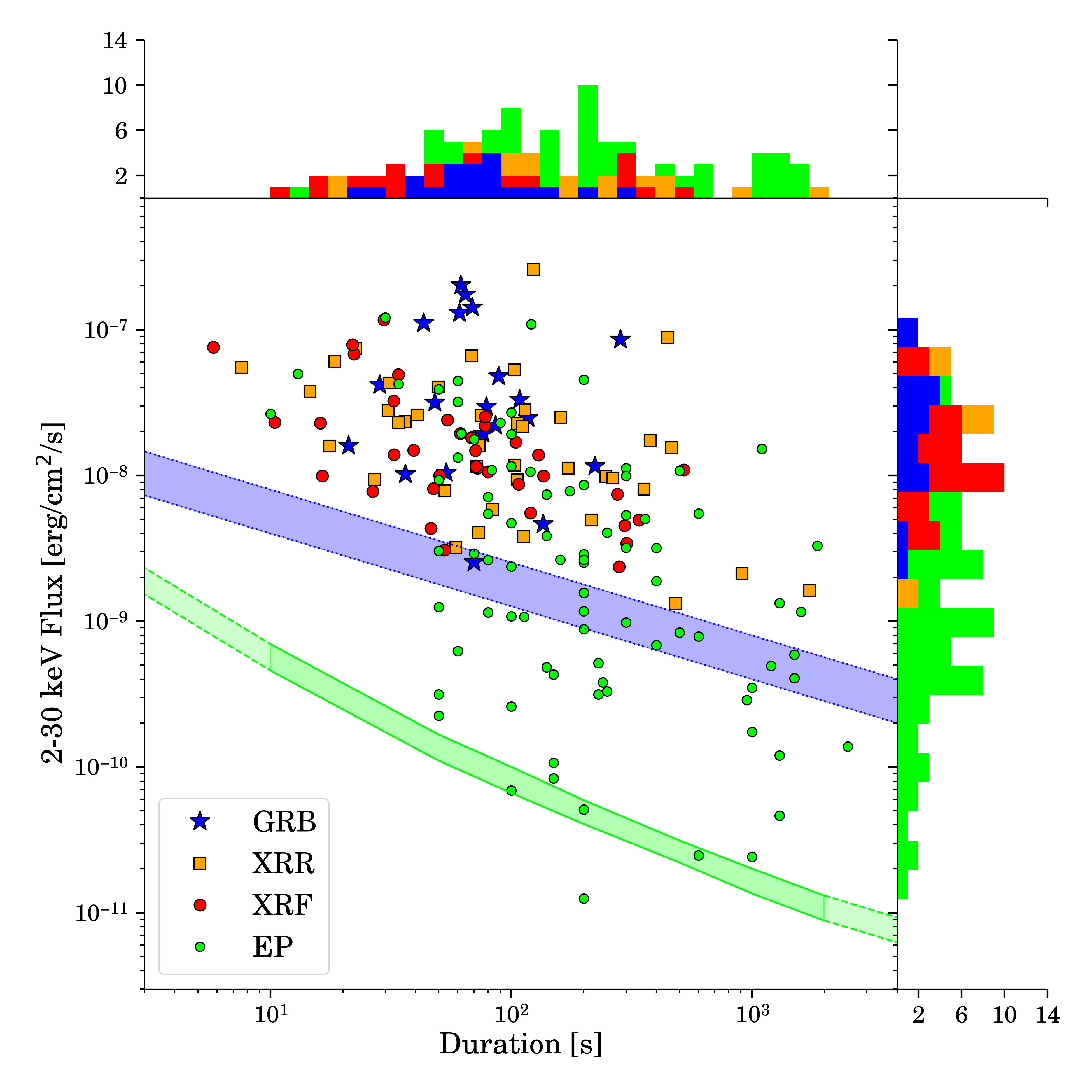}
  \caption{Flux vs $T_{90}$ for \sax GRB (blue stars), XRR (orange squares) and XRFs (red circles). EP events are represented in lime. The colors indicate the same classes also in the histograms. The \sax WFC sensitivity is represented with the blue shaded area, while EP-WXT sensitivity in lime. For a detailed explanation on how the sensitivities were estimated see Paper I.}
\label{fig:EP_SAX_flux}
\end{figure}

\begin{figure}
\centering
 \includegraphics[width=0.5\textwidth]{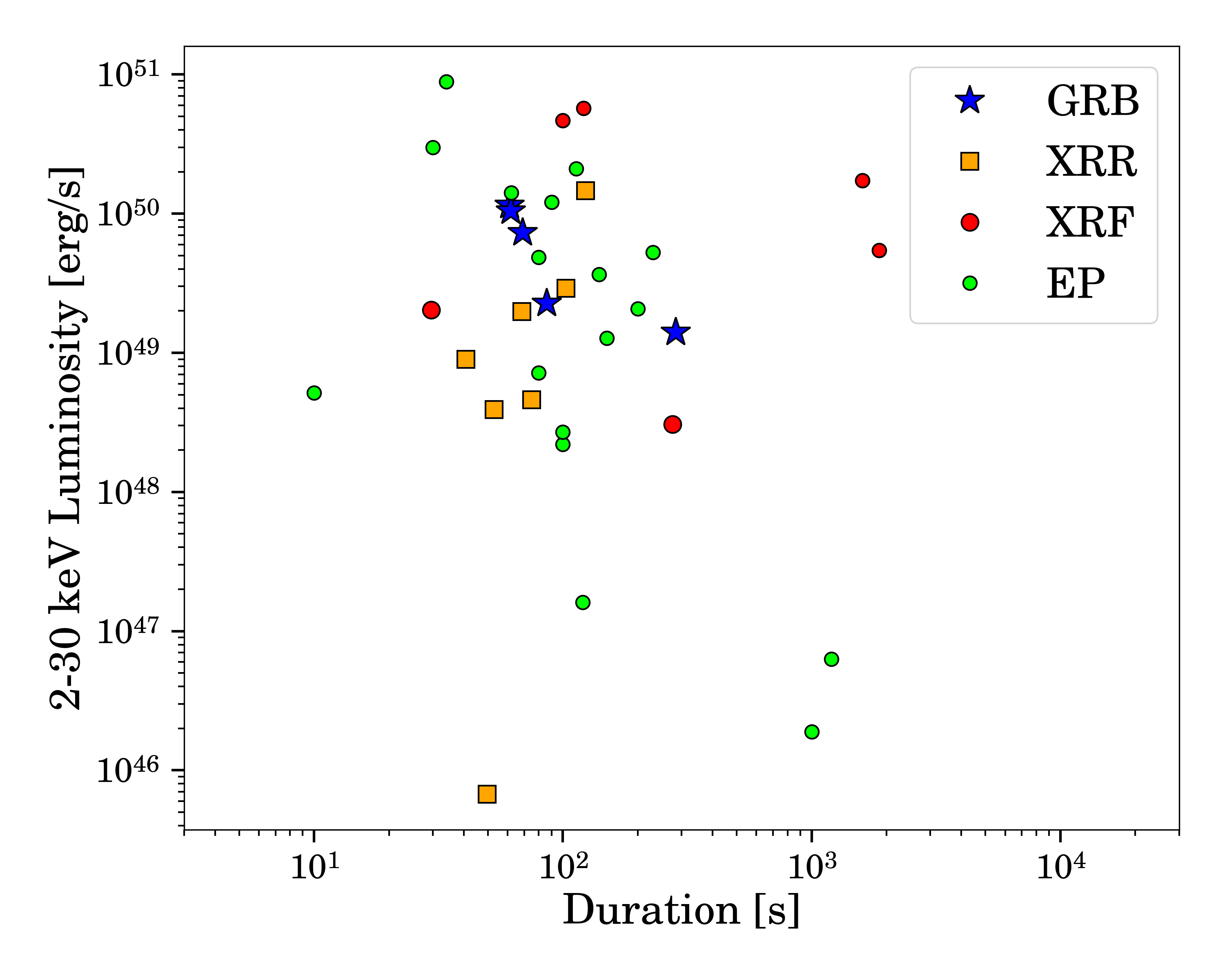}
  \caption{Rest-frame 2-30 keV luminosity as a function of $\rm T_{90}$ for \sax GRB (blue stars), XRR (orange squares) and XRFs (red circles). EP transients are represented in lime.}
\label{fig:EP_SAX_luminosity}
\end{figure}

Discovering soft X-ray transients has been quite difficult after the end of the \sax mission, due to instruments mainly observing the sky in the hard X-rays and $\gamma-$rays. However, from 2024, the number of soft FXTs is growing thanks to EP mission. EP-WXT has been discovering since 2024 hundreds of FXTs \citep{Aryan2025, Wu2025, Zhang2025}. We represent the 2-30 keV fluxes and durations of the EP events reported in \citealt{Aryan2025, Wu2025} and in GCNs (General Coordinate Network) with lime circles in Fig. \ref{fig:EP_SAX_flux}. The 2-30 keV fluxes are estimated assuming the photon index value reported in GCNs, see also Paper I for a detailed explanation. The \sax bursts are represented with colored stars. We find that, while at high fluxes the \sax and EP population superimpose, EP is clearly discovering an extension of the \sax population at low fluxes, thanks to its better sensitivity (EP sensitivity in lime, \sax sensitivity in blue). This is confirmed also looking at the luminosity of the bursts with redshift in Fig. \ref{fig:EP_SAX_luminosity}. As already pointed out, EP reaches a population of fainter and more numerous events, whose presence was first hinted at by the unique very low luminosity \sax GRB980425.

\section{Conclusions}
\label{sec:conclusioni}
We have derived the spectral and
temporal properties of the prompt emission of  the  complete sample of 96 bona-fide GRB
detected by the Wide Field Cameras (2-28 keV) of \sax,  complementing these data with the higher energy Gamma-Ray Burst Monitor (40-700 keV) simultaneous observations and with ancillary information regarding their afterglow properties. 
This corresponds to 480$\pm$50 yr$^{-1}$, having accounted for the FOV of 0.47 sr and 5.33 yr of observations (paper I). 
Our main findings are the following.

\begin{itemize}
\item We have assessed the properties
of the soft population of GRB, namely X-ray flashes (XRFs), in comparison with normal GRBs. On the
basis of the spectral shape we have found that 36 events are X-ray flashes, 40 X-ray rich
events (XRR), and 20 normal GRBs. We analyzed the distribution of the spectral parameters of
the Band function, $\alpha$, $\beta$ and $\epc$, finding that the spectral indexes of the three classes are broadly
similar. On the contrary, the peak energy was the parameter driving the spectra shape, from $\approx 9$ keV to
 10 times larger  in GRBs. 
\item  The  duration (T$_{90}$) in the X-ray range was similar
in  the three classes, clustering respectively at around 70 s. Likewise, a similar duration  of 25 s was observed in the gamma-ray range.
The longer duration in X-rays  is consistent with the well known hard-to-soft evolution. 
\item For the 64  events that are detected in both instruments we found that 9 events exhibit an X-ray precursor taking place from 14 to 105 s before the onset of the gamma-ray burst. 
\item  About 90\% of the events that were identified in real time and subsequently followed up by \sax exhibited an X-ray afterglow, with a similar fraction for the three classes. In the optical and radio the the corresponding fractions were 35\% and 33\%. 
\item Both the XRFs and XRR events are falling onto the $\rm E_{p,i}-E{_{iso}}$ relation followed by GRBs, with one exception (GRB 980425).
\item All the similarities in the spectrum, duration and afterglow properties suggest common progenitors for the three classes, where the differences are likely a combination of the effect of different baryon loading, energy, structure and orientation of the jet with respect to the observer.
\item A comparison  with \textit{Einstein Probe} showed that the latter, thanks to its sensitivity, reaches out a population of fainter and more numerous events, whose presence was firstly hinted at by the unique very low luminosity \sax GRB980425. 
\end{itemize}

\begin{acknowledgements}
We acknowledge support by the European Union Horizon 2020 programme under the AHEAD2020 project (grant agreement number 871158). This work has been also supported by ASI (Italian Space Agency) through the Contracts no. 2019-27-HH.0, and no. 2025-5-U.0.

\end{acknowledgements}

\bibliographystyle{aa} 
\bibliography{bibliography.bib}

\begin{appendix}
\onecolumn
\clearpage

\section{Data tables}

\begin{longtable}{cccccccc}
    \caption[]{Counting rates and durations. In the first column the GRB name is reported, while from the second to the fifth columns $T_{50}$ and $T_{90}$ values are reported for WFC and GRBM instruments. The WFC and GRBM mean count rates are reported in the sixth and seventh columns, while their softness ratio $\srobs$ in the last column. The WFC and GRBM mean rates are estimated in the same exposure time. 2$\sigma$ lower limits on the $\srobs$ are indicated with $>$.}
    \label{tab1}\\
    \hline
    \noalign{\smallskip}
    Event & $\rm T_{50,WFC}$ & $\rm T_{90,WFC}$ & $\rm T_{50,GRBM}$ &  $\rm T_{90,GRBM}$ & WFC mean rate & GRBM mean rate & $ \srobs$ \\
     & s & s & s & s & cts/s/cm$^2$ & cts/s & $10^{-2}$ \\
    \hline
    \noalign{\smallskip}
960720  & 9 & 21 & 1 & 6 & 0.70  $\pm$ 0.07  & 112.8 $\pm$ 2.2 & 0.62  $\pm$ 0.08  \\
960726  & 42 & 78 & - & - & 1.19  $\pm$ 0.17  & 18.1  $\pm$ 7.5 & 6.57  $\pm$ 3.62  \\
960806  & 14  & 23 & 8 & 13 & 0.54  $\pm$ 0.06  & 29.4  $\pm$ 8.9 & 1.83  $\pm$ 0.74  \\
961229  & 38  & 53 & - &  - & 0.21  $\pm$ 0.04  & 0.0 $\pm$ 1.2 & >8.8 \\
961229B & 162 & 340 & - & - & 0.26 $\pm$ 0.04 & 2.0$\pm$1.5 & 13.0$\pm$9.9 \\
970111  & 18  & 43 & 11 & 31 & 3.37  $\pm$ 0.15  & 1321.4  $\pm$ 7.5 & 0.25  $\pm$ 0.01  \\
970228  & 43  & 70 & 44 & 56 & 3.25  $\pm$ 0.16  & 151.1 $\pm$ 1.9 & 2.15  $\pm$ 0.13  \\
970402  & 75  & 102 & 72  & 105 & 0.46  $\pm$ 0.03  & 60.8  $\pm$ 1.0 & 0.76  $\pm$ 0.06  \\
970508  & 11  & 41 & 5  & 14 & 1.33  $\pm$ 0.07  & 66.9  $\pm$ 1.3 & 1.99  $\pm$ 0.14  \\
971019  & 12  & 22 & -  & - & 3.72  $\pm$ 0.31  & 12.7  $\pm$ 6.6 & 29.21 $\pm$ 17.47 \\
971019B  & 107 & 447 & - & - & 0.13 $\pm$ 0.02 & 1.6$\pm$2.0 & > 3.3 \\ 
971024  & 7 & 16 & 5  & 27 & 1.30  $\pm$ 0.07  & 29.4  $\pm$ 11.7  & 4.42  $\pm$ 2.00  \\
971206  & 18  & 76 & 15 & 41 & 0.65  $\pm$ 0.13  & 149.3 $\pm$ 6.2 & 0.43  $\pm$ 0.10  \\
971214  & 20  & 86 & 15  & 30 & 0.82  $\pm$ 0.15  & 346.1 $\pm$ 8.9 & 0.24  $\pm$ 0.05  \\
971227  & 4 & 15 & 3  & 10 & 1.25  $\pm$ 0.14  & 166.6 $\pm$ 4.2 & 0.75  $\pm$ 0.10  \\
980109  &  14 & 36  &  7  & 18 & 0.80  $\pm$ 0.06  & 99.7  $\pm$ 1.4 & 0.80  $\pm$ 0.07  \\
980128  & 68 & 112  & -  & - & 0.18  $\pm$ 0.02  & 7.6 $\pm$ 3.3 & 2.42  $\pm$ 1.24  \\
980326  & 2  & 6 &  1  &  8 & 3.45  $\pm$ 0.41  & 88.5  $\pm$ 2.4 & 3.90  $\pm$ 0.57  \\
980329  & 11  & 64 & 6  & 19 & 5.03  $\pm$ 0.49  & 2520.2  $\pm$ 0.8 & 0.20  $\pm$ 0.02  \\
980412  & 192 & 480 & - & - & 0.10 $\pm$ 0.02 & 4.8$\pm$1.8 & 2.1$\pm$0.9\\
980415  & 34 & 73 & - & - & 0.15 $\pm$ 0.03 & 0.9$\pm$2.9 & > 2.6 \\ 
980425  & 40 & 50 & 10  & 22 & 1.79  $\pm$ 0.17  & 98.3  $\pm$ 6.2 & 1.82  $\pm$ 0.29  \\
980429  & 61  & 130 & 80 & 165 & 0.82  $\pm$ 0.04  & 4.0 $\pm$ 2.4 & 20.42 $\pm$ 13.16 \\
980515  & 13  & 31 & 9  & 25 & 1.17  $\pm$ 0.11  & 100.0 $\pm$ 1.7 & 1.17  $\pm$ 0.13  \\
980519  & 45  & 161 & 8  & 28 & 1.22  $\pm$ 0.02  & 85.8  $\pm$ 3.3 & 1.43  $\pm$ 0.08  \\
980613  & 24  & 57 & 7  & 42 & 0.38  $\pm$ 0.05  & 48.1  $\pm$ 7.1 & 0.79  $\pm$ 0.22  \\
980614  & 131 & 314 & - & - & 0.20 $\pm$ 0.04 & 0.02$\pm$3.0 & > 3.3 \\
980706  & 10  & 84 & -  & - & 0.35  $\pm$ 0.05  & 16.3  $\pm$ 7.6 & 2.14  $\pm$ 1.30  \\
980824  & 5 & 46 & -  & - & 0.29  $\pm$ 0.05  & 1.0 $\pm$ 5.5 & > 2.6  \\
981018  & 183 & 356  & 77 & 115 & 0.38 $\pm$ 0.03 & 4.0$\pm$1.4 & 9.5$\pm$3.4 \\
981226  & 40 & 136 & 41 & 61 & 0.59  $\pm$ 0.03  & 12.9  $\pm$ 4.5 & 4.61  $\pm$ 1.80  \\
990123  & 29 & 61 & 40 & 61 & 4.57  $\pm$ 0.06  & 2288.8  $\pm$ 6.8 & 0.20  $\pm$ 0.01 \\
990217  & 29  & 173 & 17  & 68 & 0.51  $\pm$ 0.04  & 51.9  $\pm$ 1.4 & 0.99  $\pm$ 0.10  \\
990328  & 9 & 296 & -  & - & 0.25  $\pm$ 0.03  & 3.7 $\pm$ 4.6 & > 2.7 \\ 
990413  & 1000 & 1735 & 200 & 350 & 0.068 $\pm$ 0.006 & 2.4$\pm$1.0 & 2.83$\pm$1.21 \\
990510  & 49  & 100 & 6  & 57 & 2.23  $\pm$ 0.08  & 218.2 $\pm$ 1.4 & 1.02  $\pm$ 0.04  \\
990520  & 6 & 10 &-  & - & 1.19  $\pm$ 0.11  & 21.8  $\pm$ 9.1 & 5.45  $\pm$ 2.76  \\
990526  & 12  & 27 & -  & - & 0.55  $\pm$ 0.07  & 3.3 $\pm$ 5.1 & > 5.4 \\ 
990625  & 8  & 36 & 4  & 10 & 1.50  $\pm$ 0.13  & 45.1  $\pm$ 1.4 & 3.33  $\pm$ 0.40  \\
990627  & 15  & 54 & 16  & 25 & 0.43  $\pm$ 0.03  & 80.2  $\pm$ 1.4 & 0.53  $\pm$ 0.05  \\
990704  & 11  & 22 & 5  & 22 & 4.41  $\pm$ 0.16  & 40.8  $\pm$ 1.3 & 10.80 $\pm$ 0.72  \\
990705  & 20  & 69 & 15  & 32 & 4.78  $\pm$ 0.25  & 1869.8  $\pm$ 6.3 & 0.26  $\pm$ 0.01  \\
990712  & 10  & 31 & 13  & 19 & 6.34  $\pm$ 0.15  & 205.3 $\pm$ 2.3 & 3.09  $\pm$ 0.11  \\
990806  & 18  & 34 & 9  & 11 & 0.97  $\pm$ 0.07  & 84.7  $\pm$ 1.5 & 1.15  $\pm$ 0.10  \\
990907  & 88  & 223 & 58  & 145 & 0.43  $\pm$ 0.02  & 72.9  $\pm$ 3.3 & 0.60  $\pm$ 0.06  \\
990908  & 39  & 101 & 31  & 47 & 0.86  $\pm$ 0.02  & 22.0  $\pm$ 3.6 & 3.90  $\pm$ 0.73  \\
991014  & 3  & 8 & 1.2  & 2.3 & 2.12  $\pm$ 0.29  & 219.7 $\pm$ 5.6 & 0.96  $\pm$ 0.16  \\
991026  & 60  & 100 & 33  & 57 & 1.19  $\pm$ 0.04  & 50.1  $\pm$ 4.2 & 2.38  $\pm$ 0.27  \\
991030  & 30  & 110 & 19  & 30 & 1.08  $\pm$ 0.10  & 109.2 $\pm$ 6.1 & 0.98  $\pm$ 0.15  \\
991105  & 36  & 92 & 29 & 61 & 0.93  $\pm$ 0.08  & 66.8  $\pm$ 5.6 & 1.39  $\pm$ 0.23  \\
991106  & 8  & 33 & 8 & 14 & 0.72  $\pm$ 0.08  & 18.2  $\pm$ 10.8  & 3.93  $\pm$ 2.80  \\
991217  & 14  & 107 & -  & - & 0.51  $\pm$ 0.03  & 5.1 $\pm$ 4.1 & 10.03 $\pm$ 8.64  \\
000206  & 16  & 39 & 9  & 26 & 0.89  $\pm$ 0.05  & 17.3  $\pm$ 7.2 & 5.12  $\pm$ 2.41  \\
000208  & 16  & 59 & 38  & 48 & 0.14  $\pm$ 0.02  & 7.4 $\pm$ 3.4 & 1.94  $\pm$ 1.19  \\
000210  & 10  & 62 & 3  & 9 & 7.45  $\pm$ 0.35  & 3064.4  $\pm$ 14.2  & 0.24  $\pm$ 0.01  \\
000214  & 59  & 118 & 3  & 8 & 1.03  $\pm$ 0.04  & 189.7 $\pm$ 4.6 & 0.54  $\pm$ 0.03  \\
000218  & 172 & 377 & 60  & 252 & 0.76  $\pm$ 0.05  & 53.0  $\pm$ 3.8 & 1.43  $\pm$ 0.20  \\
000416  & 29  & 68 & -  & - & 1.25  $\pm$ 0.09  & 4.6 $\pm$ 4.4 & > 1.4 \\ 
000424  & 10  & 34 &  - & - & 1.72  $\pm$ 0.18  & 1.1$\pm$8.4 & > 1.0 \\ 
000528  & 30  & 109 & 27  & 65 & 1.06  $\pm$ 0.03  & 239.3 $\pm$ 4.0 & 0.44  $\pm$ 0.02  \\
000529  & 18  & 31 & 3  & 13 & 1.51  $\pm$ 0.12  & 262.8 $\pm$ 9.4 & 0.57  $\pm$ 0.07  \\
000608  & 23  & 48 & -  & - & 0.47  $\pm$ 0.07  & 13.5  $\pm$ 4.6 & 3.46  $\pm$ 1.67  \\
000615  & 30  & 78 & 4  & 13 & 1.55  $\pm$ 0.11  & 17.5  $\pm$ 3.7 & 8.83  $\pm$ 2.50  \\
000620  & 6 & 31 & 11  & 25 & 1.62  $\pm$ 0.27  & 148.3 $\pm$ 9.4 & 1.09  $\pm$ 0.25  \\
000628  & 13 & 80 & -  & -  & 0.59  $\pm$ 0.07  & 11.0  $\pm$ 3.8 & 5.41  $\pm$ 2.49  \\
000630  & 32  & 72 & 13  & 26 & 0.75  $\pm$ 0.06  & 16.7  $\pm$ 4.8 & 4.48  $\pm$ 1.67  \\
000901  & 50 & 100 & - & - & 0.28 $\pm$ 0.07 & 0.4$\pm$2.5 & > 5.6 \\ 
001011  & 23  & 48 & 8  & 24 & 1.19  $\pm$ 0.03  & 441.4 $\pm$ 5.8 & 0.27  $\pm$ 0.01  \\
001024  & 14  & 18 & - & - & 0.72  $\pm$ 0.11  & 35.1  $\pm$ 12.5  & 2.05  $\pm$ 1.04  \\
001028  & 63 & 70 & - & - & 0.09 $\pm$ 0.03 & 2.0$\pm$2.6 & > 1.7 \\ 
001101  & 46  & 120 & -  & - & 0.35  $\pm$ 0.02  & -3.1  $\pm$ 3.0 & > 5.8 \\ 
001109  & 39  & 75 & 32  & 54 & 1.12  $\pm$ 0.06  & 98.5  $\pm$ 5.1 & 1.14  $\pm$ 0.12  \\
001110  & 16  & 61 & -  & - & 0.93  $\pm$ 0.09  & 4.0 $\pm$ 4.8 & > 9.7 \\ 
010213  & 9  & 22 & 6  & 10 & 2.74  $\pm$ 0.16  & 301.7 $\pm$ 8.5 & 0.91  $\pm$ 0.08  \\
010214  & 24  & 79 & 7  & 17 & 0.97  $\pm$ 0.06  & 217.3 $\pm$ 8.8 & 0.45  $\pm$ 0.05  \\
010219  & 14  & 33 & 13 & 17 & 2.30  $\pm$ 0.20  & 8.4 $\pm$ 5.8 & 27.48 $\pm$ 21.42 \\
010220  & 22  & 247 & 16  & 31 & 0.32  $\pm$ 0.02  & 40.4  $\pm$ 3.3 & 0.79  $\pm$ 0.11  \\
010222  & 41  & 124 & 14 & 74 & 11.96 $\pm$ 0.19  & 1489.0  $\pm$ 6.8 & 0.80  $\pm$ 0.02  \\
010304  & 12  & 19 & 4  & 12 & 1.98  $\pm$ 0.11  & 272.4 $\pm$ 8.9 & 0.73  $\pm$ 0.06  \\
010320  & 33  & 71 & -  & - & 0.95  $\pm$ 0.11  & -4.7  $\pm$ 5.1 & > 9.3 \\  
010324  & 202 & 465 & 191 & 292 & 0.75  $\pm$ 0.08  & 47.8  $\pm$ 2.4 & 1.57  $\pm$ 0.25  \\
010412  & 49  & 89 & 24   & 60 & 1.81  $\pm$ 0.07  & 466.6 $\pm$ 4.6 & 0.39  $\pm$ 0.02  \\
010501A & 86  & 215 & 44  & 66 & 0.22  $\pm$ 0.02  & 18.4  $\pm$ 4.5 & 1.19  $\pm$ 0.37  \\
010501B & 72  & 72 & 23  & 30 & 0.28  $\pm$ 0.04  & 58.0  $\pm$ 4.9 & 0.48  $\pm$ 0.10  \\
010518  & 110 & 264 & 12  & 22 & 0.38  $\pm$ 0.05  & 52.7  $\pm$ 5.0 & 0.72  $\pm$ 0.17  \\
010527  & 117 & 313 & 14 & 26 & 0.72  $\pm$ 0.09  & 15.7  $\pm$ 4.8 & 4.59  $\pm$ 1.98  \\
010707  & 10  & 16 & - & - & 0.72  $\pm$ 0.05  & 1.9 $\pm$ 6.1 & > 5.9 \\ 
011030  & 133 & 280 & -  & - & 0.14  $\pm$ 0.01  & 2.5 $\pm$ 2.2 & 5.39  $\pm$ 5.00  \\
011121  & 20  & 284 & 14  & 47 & 3.65  $\pm$ 0.06  & 755.3 $\pm$ 3.5 & 0.48  $\pm$ 0.01  \\
011211  & 27  & 276 & 27 & 51 & 0.37  $\pm$ 0.02  & 8.8 $\pm$ 2.4 & 4.21  $\pm$ 1.31  \\
020113  & 401 & 909 & 405 & 732 & 0.09 $\pm$ 0.01 & 4.1$\pm$1.6 & 2.20$\pm$0.98\\
020118  & 17  & 54 & -  & - & 1.09  $\pm$ 0.09  & 14.9  $\pm$ 5.8 & 7.34  $\pm$ 3.45  \\
020321  & 100 & 136 & 24  & 51 &  0.21  $\pm$ 0.03  & 48.4  $\pm$ 5.2 & 0.44  $\pm$ 0.10  \\
020322  & 37  & 106 & 4  & 11 &  0.45  $\pm$ 0.04  & 20.5  $\pm$ 5.2 & 2.21  $\pm$ 0.75  \\
020409  & 38  & 74 & 20  & 36 &  0.64  $\pm$ 0.11  & 42.5  $\pm$ 5.4 & 1.50  $\pm$ 0.46  \\
020427  & 32  & 50 & -  & - &  0.62  $\pm$ 0.04  & 8.9 $\pm$ 3.7 & 6.96  $\pm$ 3.39  \\
    \noalign{\smallskip}
    \hline
    
\end{longtable}

\clearpage
\twocolumn

\clearpage
\onecolumn

\begin{longtable}{ccccccccccccc}
    \caption[]{Spectral properties of individual bursts. For each burst (first column) spectrum, we report the exposure time (second column), the fluence in the 2-30 keV and 30-400 keV (third and forth columns), their softness ratio SR (fifth column). The slope $\rm \Gamma_{PL}$ of the power law fit and its reduced $\cq$ are written in the sixth and seventh columns. The results of the Band function fit are the slopes $\alpha$ and $\beta$ in columns eight and nine, the peak energy in column ten, the reduced $\cq$ in column eleven. The $\Delta \chi^2$ between the power law and the Band model is reported in column twelve, and the degree of freedom of the Band model in the last column.  }
    \label{tab:spectra}\\
    \hline
    \noalign{\smallskip}
    GRB & $\rm T_{exp}$ & S(2-30) & S(30-400) & $\sr$ & $\rm \Gamma_{PL}$ & $\cnu$ & $\alpha$ & $\beta$ & $\epc$ & $\cnu$ & $\Delta \chi^2$ & $\nu$ \\
     & s & $\fluence$ & $\fluence$ &  &  &  & & & keV & &  \\
     \noalign{\smallskip}
    \hline
    \noalign{\smallskip}
960720  &   23  &   3.67E-07    &   1.28E-06    &   0.29    &   -1.51   &   1.07    &   -1.60   $^{+    0.11    }_{-    0.04    }$  &   -2.20  &   15000   $^{+    100 }_{-    14800   }$  &   1.1 &   0.26    &   27   \\
960726  &   25  &   5.48E-07    &   2.92E-07    &   1.88    &   -2.1    &   1.2 &   -1.00   &   -2.47   $^{+    0.3 }_{-    100 }$  &   13  $^{+    17  }_{-    6   }$  &   0.98    &   7.14    &   27   \\
960806  &   34  &   3.20E-07    &   4.68E-07    &   0.68    &   -1.83   &   0.71    &   -1.80   $^{+    0.38    }_{-    0.16    }$  &   -2.20   &   300 $^{+    10000   }_{-    300 }$  &   0.73    &   0.15    &   28   \\
961229  &   600 &   1.86E-06    &   5.90E-08    &   31.43   &   -2.33   &   0.66    &   -1.00  &   -9.95   $^{+    7.5 }_{-    100 }$  &   8   $^{+    11  }_{-    4   }$  &   0.47    &   6.29    &   26  \\
961229b & 380 & 1.878E-6 & 1.220E-6 & 1.54 & -1.66 & 0.64 & -1.00 & -1.69$\pm$1.22 & 11.9$\pm$9.2 & 0.71 & -1.96 & 28 \\
970111  &   60  &   6.66E-06    &   4.13E-05    &   0.16    &   -1.25   &   34.7    &   -2.53   $^{+    0.13    }_{-    0.05    }$  &   -9.9    $^{+    9.4 }_{-    100 }$  &   123 $^{+    5   }_{-    25  }$  &   0.79    &   951.06  &   26   \\
970228  &   47  &   3.10E-06    &   4.14E-06    &   0.75    &   -1.83   &   1.37    &   -1.50   $^{+    1.42    }_{-    0.17    }$  &   -9.47   $^{+    7.0 }_{-    100 }$  &   65  $^{+    29  }_{-    57  }$  &   1.05    &   11.06   &   26   \\
970402  &   120 &   1.41E-06    &   3.87E-06    &   0.36    &   -1.51   &   1.36    &   -0.89   $^{+    1.44    }_{-    0.29    }$  &   -1.83   $^{+    0.4 }_{-    100 }$  &   45  $^{+    81  }_{-    45  }$  &   0.71    &   19.62   &   26   \\
970508  &   27  &   7.00E-07    &   1.05E-06    &   0.67    &   -1.84   &   0.68    &   -1.82   $^{+    0.16    }_{-    0.04    }$  &   -2.20   &   150 $^{+    10000   }_{-    210 }$  &   0.67    &   0.94    &   26   \\
971019  &   23  &   1.57E-06    &   2.66E-07    &   5.88    &   -2.33   &   2.87    &   -0.47   $^{+    1.15    }_{-    0.71    }$  &   -3.44   $^{+    0.6 }_{-    100 }$  &   12  $^{+    8   }_{-    4   }$  &   1.33    &   44.24   &   25   \\
971019b & 200 & 1.768E-5 & 3.035E-5 & 0.58 & -1.36 & 1.32 & -1.06$\pm$0.44 & -8.16$\pm$1.00 & 22$\pm$33 & 1.35 & -0.81 & 27 \\
971024  &   13  &   2.99E-07    &   1.97E-07    &   1.51    &   -1.95   &   1.67    &   -1.00  &   -2.26   $^{+    0.2 }_{-    0.48    }$  &   10  $^{+    5   }_{-    3   }$  &   0.93    &   22.39   &   28   \\
971206  &   51  &   9.84E-07    &   3.84E-06    &   0.26    &   -1.43   &   1.09    &   -1.05   $^{+    0.39    }_{-    0.27    }$  &   -2.20   &   120 $^{+    240 }_{-    58  }$  &   0.89    &   6.69    &   28   \\
971214  &   30  &   6.59E-07    &   5.13E-06    &   0.13    &   -1.15   &   1.29    &   -0.78   $^{+    0.21    }_{-    0.21    }$  &   -2.20  &   167 $^{+    190 }_{-    70  }$  &   0.83    &   15  &   27   \\
971227  &   10  &   3.77E-07    &   9.14E-07    &   0.41    &   -1.43   &   1.53    &   0.20    $^{+    0.2 }_{-    0.86    }$  &   -1.96   $^{+    0.3 }_{-    100 }$  &   27  $^{+    47  }_{-    15  }$  &   0.83    &   20.56   &   25   \\
980109  &   50  &   5.09E-07    &   2.76E-06    &   0.18    &   -1.52   &   1.28    &   -0.93   $^{+    0.36    }_{-    0.28    }$  &   -5.17   $^{+    4.2 }_{-    100 }$  &   91  $^{+    65  }_{-    91  }$  &   0.98    &   10.06   &   25   \\
980128  &   80  &   3.04E-07    &   4.18E-07    &   0.73    &   -1.76   &   1.35    &   -0.78   $^{+    2.18    }_{-    0.61    }$  &   -2.20 &   17  $^{+    26  }_{-    9   }$  &   1   &   10.8    &   27   \\
980326  &   9   &   6.86E-07    &   5.99E-07    &   1.15    &   -1.92   &   1.61    &   -0.93   $^{+    2   }_{-    0.46    }$  &   -4.07   $^{+    2.0 }_{-    100 }$  &   35  $^{+    9   }_{-    24  }$  &   1.28    &   11.8    &   26   \\
980329  &   25  &   4.35E-06    &   3.18E-05    &   0.14    &   -1.11   &   1.52    &   -0.56   $^{+    0.33    }_{-    0.31    }$  &   -2.20 &   124 $^{+    140 }_{-    43  }$  &   1.2 &   9.84    &   26   \\
980412  & 300 & 3.976E-07 & 4.146E-7 & 0.96 & -1.97 & 0.85 & -1.0 & -1.71$\pm$1.51 & 11$\pm$11 & 0.95 & -2.8 & 28 \\
980415  & 100 & 4.096E-7 & 5.172E-6 & 0.08 & -1.01 & 1.20 & -1.0 & -2.0$\pm$1.0 & 9999$^{+100}_{-8900}$ & 1.24 & -1.12 & 28 \\
980425  &   55  &   2.23E-06    &   2.86E-06    &   0.78    &   -1.84   &   1.13    &   -1.39   $^{+    0.44    }_{-    0.3 }$  &   -2.20 &   40  $^{+    55  }_{-    17  }$  &   0.92    &   7.72    &   26   \\
980429  &   160 &   2.20E-06    &   5.57E-07    &   3.95    &   -2.32   &   1.99    &   -1.75   $^{+    0.28    }_{-    0.25    }$  &   -10 $^{+    8.3 }_{-    100 }$  &   10  $^{+    6   }_{-    7   }$  &   1.38    &   19.84   &   26   \\
980515  &   40  &   1.11E-06    &   2.22E-06    &   0.50    &   -1.59   &   1.4 &   -0.67   $^{+    0.65    }_{-    0.48    }$  &   -2.20   &   32  $^{+    29  }_{-    10  }$  &   0.97    &   12.58   &   26   \\
980519  &   165 &   4.12E-06    &   7.42E-06    &   0.56    &   -1.75   &   1.04    &   -1.63   $^{+    0.06    }_{-    0.06    }$  &   -2.20 &   146 $^{+    130 }_{-    40  }$  &   0.83    &   6.71    &   27   \\
980613  &   37  &   2.89E-07    &   8.86E-07    &   0.33    &   -1.55   &   0.78    &   -1.55   $^{+    0.5 }_{-    0.1 }$  &   -2.20 &   9999    $^{+    100 }_{-    9040    }$  &   0.81    &   -0.52   &   26   \\
980614  & 350 & 1.20E-06 & 8.52E-07 & 1.41 & -0.96 & 0.78 & -1.00 & -2.23$^{+0.5}_{-0.3}$ & 13$^{+10}_{-5}$ & 0.89 & -0.84 & 28 \\
980706  &   30  &   1.76E-07    &   1.96E-07    &   0.90    &   -1.94   &   1.02    &   -1.00  &   -1.94   $^{+    0.2 }_{-    0.25    }$  &   2   $^{+    29  }_{-    2   }$  &   1.05    &   0.21    &   27   \\
980824  &   29  &   1.26E-07    &   6.30E-08    &   1.99    &   -2.14   &   0.71    &   -1.00  &   -2.27   $^{+    0.4 }_{-    100 }$  &   4   $^{+    6   }_{-    4   }$  &   0.7 &   0.98    &   27   \\
981018  & 420 & 3.383E-6 & 9.987E-6 & 0.34 & -1.57 & 1.01 & -1.00 & -8.9$\pm$1.0 & 11$\pm$3 & 1.14 & 3.64 & 28 \\
981226  &   80  &   7.94E-07    &   6.23E-07    &   1.27    &   -2.02   &   0.93    &   -1.00  &   -2.1    $^{+    0.1 }_{-    0.17    }$  &   5   $^{+    2   }_{-    5   }$  &   0.9 &   1.77    &   28   \\
990123  &   80  &   1.04E-05    &   8.47E-05    &   0.12    &   -1.13   &   8.09    &   -0.59   $^{+    0.2 }_{-    0.11    }$  &   -1.23   $^{+    0.0 }_{-    0.04    }$  &   42  $^{+    19  }_{-    15  }$  &   1.28    &   193.24  &   26   \\
990217  &   30  &   3.35E-07    &   7.90E-07    &   0.42    &   -1.66   &   0.72    &   -1.66   $^{+    0.08    }_{-    0.03    }$  &   -2.20  &   10000   $^{+    100 }_{-    9700    }$  &   0.74    &   5.2 &   26   \\
990328  &   40  &   1.81E-07    &   4.63E-08    &   3.91    &   -2.04   &   0.61    &   -1.00   &   -2.88   $^{+    1.7 }_{-    100 }$  &   10  $^{+    11  }_{-    5   }$  &   0.49    &   3.97    &   28   \\
990413  & 836 & 1.358E-6 & 1.801E-6 & 0.75 & -1.00 & 0.59 & -1.00 & -8.45$\pm$1.00 & 9999$^{+100}_{-9038}$ & 0.60 & -0.28 & 28 \\
990510  &   120 &   6.35E-06    &   1.44E-05    &   0.44    &   -1.63   &   1.52    &   -1.30   $^{+    0.13    }_{-    0.13    }$  &   -3.51   $^{+    2.2 }_{-    100 }$  &   103 $^{+    40  }_{-    100 }$  &   0.82    &   20.54   &   25   \\
990520  &   10  &   2.32E-07    &   1.30E-07    &   1.78    &   -2.02   &   0.79    &   -1.00  &   -2.57   $^{+    0.4 }_{-    100 }$  &   16  $^{+    19  }_{-    8   }$  &   0.45    &   10.31   &   28   \\
990526  &   31  &   2.40E-07    &   1.03E-07    &   2.33    &   -2.32   &   1.15    &   -1.00  &   -2.32   $^{+    0.3 }_{-    2.15    }$  &   3   $^{+    3   }_{-    3   }$  &   1.1 &   2.5 &   27   \\
990625  &   23  &   5.38E-07    &   5.69E-07    &   0.95    &   -1.96   &   0.89    &   -1.00  &   -1.96   $^{+    0.1 }_{-    0.04    }$  &   1   $^{+    8   }_{-    1   }$  &   0.93    &   -3.90   &   26   \\
990627  &   40  &   4.14E-07    &   1.66E-06    &   0.25    &   -1.41   &   0.88    &   -1.09   $^{+    0.27    }_{-    0.29    }$  &   -2.20 &   136 $^{+    1090    }_{-    58  }$  &   0.78    &   3.48    &   26   \\
990704  &   32  &   2.52E-06    &   9.02E-07    &   2.79    &   -2.28   &   4.84    &   -1.15   $^{+    0.65    }_{-    0.36    }$  &   -2.73   $^{+    0.2 }_{-    100 }$  &   12  $^{+    10  }_{-    4   }$  &   1.02    &   109 &   26   \\
990705  &   50  &   7.12E-06    &   4.86E-05    &   0.15    &   -1.2    &   1.98    &   -0.86   $^{+    0.16    }_{-    0.16    }$  &   -4.42   $^{+    3.2 }_{-    100 }$  &   198 $^{+    150 }_{-    198 }$  &   1.64    &   12.46   &   25   \\
990712  &   40  &   4.69E-06    &   4.59E-06    &   1.02    &   -1.98   &   2.32    &   -1.00   &   -2.03   $^{+    0.0 }_{-    0.03    }$  &   27.8   $^{+    1.3   }_{-    1.0   }$  &   0.99    &   36.9    &   26   \\
990806  &   35  &   8.04E-07    &   1.08E-06    &   0.74    &   -1.67   &   0.86    &   -1.19   $^{+    0.32    }_{-    0.32    }$  &   -2.20 &   36  $^{+    28  }_{-    10  }$  &   0.61    &   7.61    &   27   \\
990907  &   170 &   1.98E-06    &   6.36E-06    &   0.31    &   -1.47   &   2.15    &   -0.92   $^{+    0.24    }_{-    0.2 }$  &   -2.20 &   74  $^{+    39  }_{-    21  }$  &   1.2 &   27.8    &   27   \\
990908  &   130 &   2.20E-06    &   1.62E-06    &   1.35    &   -1.92   &   3.88    &   -1.41   $^{+    0.23    }_{-    0.2 }$  &   -2.35   $^{+    0.2 }_{-    100 }$  &   19  $^{+    23  }_{-    6   }$  &   1.37    &   73.02   &   26   \\
991014  &   7   &   3.85E-07    &   8.50E-07    &   0.45    &   -1.46   &   1.09    &   0.01    $^{+    1.1 }_{-    0.75    }$  &   -2.20  &   27  $^{+    22  }_{-    8   }$  &   0.6 &   13.83   &   26   \\
991026  &   100 &   2.27E-06    &   2.38E-06    &   0.96    &   -1.93   &   2.72    &   -0.83   $^{+    1.02    }_{-    0.6 }$  &   -2.20   $^{+    0.1 }_{-    0.06    }$  &   7   $^{+    3   }_{-    2   }$  &   2.37    &   14.54   &   26   \\
991030  &   40  &   1.12E-06    &   2.42E-06    &   0.46    &   -1.61   &   1.02    &   -1.00   &   -2.05   $^{+    0.4 }_{-    0.76    }$  &   48  $^{+    23  }_{-    20  }$  &   0.56    &   13.44   &   27   \\
991105  &   60  &   1.30E-06    &   2.22E-06    &   0.59    &   -1.67   &   1.18    &   -0.84   $^{+    0.59    }_{-    0.42    }$  &   -2.20   &   27  $^{+    25  }_{-    10  }$  &   0.68    &   14.18   &   26   \\
991106  &   7   &   9.67E-08    &   5.62E-08    &   1.72    &   -1.04   &   1.38    &   -1.00   &   -2.41   $^{+    0.4 }_{-    100 }$  &   13  $^{+    12  }_{-    5   }$  &   1.07    &   10.06   &   28   \\
991217  &   47  &   4.10E-07    &   1.67E-07    &   2.46    &   -2  &   1.88    &   -0.37   $^{+    1.26    }_{-    0.78    }$  &   -2.49   $^{+    0.3 }_{-    100 }$  &   8   $^{+    3   }_{-    2   }$  &   1.03    &   25.86   &   26   \\
000206  &   30  &   4.48E-07    &   2.91E-07    &   1.54    &   -2  &   0.99    &   -0.87   $^{+    1.54    }_{-    0.95    }$  &   -2.21   $^{+    0.2 }_{-    2.29    }$  &   7   $^{+    34  }_{-    2   }$  &   0.6 &   12.51   &   27   \\
000208  &   66  &   2.11E-07    &   2.90E-07    &   0.73    &   -1.85   &   0.65    &   -1.76   $^{+    1.5 }_{-    0.25    }$  &   -1.92   $^{+    0.2 }_{-    0.46    }$  &   100 $^{+    10000   }_{-    90  }$  &   0.66    &   0.37    &   28   \\
000210  &   21  &   4.24E-06    &   3.00E-05    &   0.14    &   -1.19   &   1.05    &   -0.65   $^{+    0.82    }_{-    0.35    }$  &   -1.27   $^{+    0.1 }_{-    1.27    }$  &   33  $^{+    290 }_{-    22  }$  &   0.59    &   14.06   &   26   \\
000214  &   85  &   2.11E-06    &   7.79E-06    &   0.27    &   -1.48   &   2.09    &   -1.48   $^{+    0.02    }_{-    0.02    }$  &   -2.20  &   10000   $^{+    100 }_{-    9500    }$  &   2.19    &   -0.61   &   27   \\
000218  &   125 &   2.16E-06    &   3.67E-06    &   0.59    &   -1.72   &   1.21    &   -1.15   $^{+    3.29    }_{-    0.32    }$  &   -2.02   $^{+    0.3 }_{-    100 }$  &   37  $^{+    72  }_{-    31  }$  &   0.86    &   11.52   &   26   \\
000416  &   45  &   8.20E-07    &   1.66E-07    &   4.95    &   -2.48   &   0.94    &   -1.00   &   -2.62   $^{+    0.2 }_{-    0.83    }$  &   3   $^{+    1   }_{-    3   }$  &   0.89    &   2.29    &   27   \\
000424  & 12 & 5.892E-7 & 4.188E-7 & 1.41 & -1.0 & 0.73 & -0.88$\pm$0.56 & -9.19$\pm$1.00 & 50$^{+  180 }_{-  40  }$ & 0.76 & -0.84 &  28 \\
000528  &   110 &   3.63E-06    &   1.43E-05    &   0.25    &   -1.3    &   11.9    &   -0.78   $^{+    0.2 }_{-    0.2 }$  &   -2.20  &   80  $^{+    23  }_{-    13  }$  &   1.07    &   294.55  &   25   \\
000529  &   21  &   8.77E-07    &   2.95E-06    &   0.30    &   -1.3    &   1.29    &   -0.72   $^{+    0.34    }_{-    0.39    }$  &   -2.20   &   63  $^{+    20  }_{-    56  }$  &   1.02    &   8.31    &   26   \\
000608  &   43  &   3.50E-07    &   2.49E-07    &   1.41    &   -2.03   &   0.66    &   -1.00   &   -2.11   $^{+    0.2 }_{-    0.31    }$  &   2   $^{+    3   }_{-    2   }$  &   0.86    &   -4.74   &   27   \\
000615  &   63  &   1.59E-06    &   7.81E-07    &   2.04    &   -2.23   &   0.7 &   -1.00   &   -2.27   $^{+    0.1 }_{-    0.13    }$  &   4   $^{+    2   }_{-    4   }$  &   0.68    &   1.24    &   27  \\
000620  &   21  &   8.99E-07    &   1.71E-06    &   0.52    &   -1.61   &   1.33    &   -0.76   $^{+    0.8 }_{-    0.6 }$  &   -2.20  &   40  $^{+    80  }_{-    16  }$  &   1.34    &   2.41    &   25   \\
000628  &   60  &   6.38E-07    &   4.71E-07    &   1.36    &   -2.03   &   0.68    &   -1.00    &   -2.20   $^{+    0.2 }_{-    0.5 }$  &   10  $^{+    12  }_{-    5   }$  &   0.54    &   4.46    &   27   \\
000630  &   70  &   7.86E-07    &   5.79E-07    &   1.36    &   -2.07   &   0.86    &   -1.00   &   -2.12   $^{+    0.1 }_{-    0.19    }$  &   5   $^{+    4   }_{-    5   }$  &   0.86    &   0.86    &   28   \\
000901  & 125 & 1.366E-6 & 1.351E-6 & 1.01 & -0.68 & 1.18 & -1.00 & -2.0$\pm$1.0 & 9999$^{+100}_{-8040}$ & 1.19 & -0.28 & 28 \\
001011  &   62  &   1.95E-06    &   1.31E-05    &   0.15    &   -1.27   &   1.48    &   -1.15   $^{+    0.07    }_{-    0.07    }$  &   -2.20  &   553 $^{+    650 }_{-    200 }$  &   1.22    &   8.5 &   27   \\
001024  &   11  &   1.74E-07    &   2.32E-07    &   0.75    &   -1.82   &   1.38    &   -1.00   &   -10 $^{+    9.0 }_{-    100 }$  &   50  $^{+    16  }_{-    39  }$  &   1.25    &   4.89    &   27   \\
001028  & 120 & 3.053E-7 & 2.626E-6 & 0.12* & -1.16 & 0.90 & -1.00 & -2.0$\pm$1.0 & 180$^{+4466}_{-100}$ & 0.90 & 0.0 & 28 \\
001101  &   84  &   4.65E-07    &   1.41E-07    &   3.29    &   -2.54   &   0.75    &   -1.00   &   -2.43   $^{+    0.2 }_{-    0.26    }$  &   1   $^{+    2   }_{-    1   }$  &   0.67    &   2.99    &   28   \\
001109  &   65  &   1.68E-06    &   3.46E-06    &   0.48    &   -1.63   &   1.58    &   -1.07   $^{+    1.04    }_{-    0.24    }$  &   -1.89   $^{+    0.2 }_{-    0.47    }$  &   36  $^{+    48  }_{-    27  }$  &   0.87    &   22.33   &   27   \\
001110  &   35  &   6.77E-07    &   2.02E-07    &   3.35    &   -2.17   &   1.66    &   -0.96   $^{+    0.75    }_{-    0.39    }$  &   -9.44   $^{+    7.4 }_{-    100 }$  &   19  $^{+    7   }_{-    9   }$  &   0.74    &   27.24   &   26   \\
010213  &   26  &   1.94E-06    &   4.50E-06    &   0.43    &   -1.56   &   2.99    &   -0.80   $^{+    0.26    }_{-    0.23    }$  &   -2.28   $^{+    0.3 }_{-    0.46    }$  &   54  $^{+    25  }_{-    16  }$  &   1.03    &   56.94   &   26   \\
010214  &   25  &   7.39E-07    &   2.91E-06    &   0.25    &   -1.37   &   1.74    &   -0.69   $^{+    0.3 }_{-    0.26    }$  &   -2.20   $^{+    0.5 }_{-    100 }$  &   81  $^{+    54  }_{-    38  }$  &   0.56    &   34.16   &   26   \\
010219  &   27  &   8.72E-07    &   2.04E-07    &   4.26    &   -2.44   &   0.84    &   -1.00   &   -2.57   $^{+    0.2 }_{-    0.43    }$  &   3   $^{+    1   }_{-    1   }$  &   0.77    &   2.8 &   28   \\
010220  &   159 &   1.57E-06    &   3.82E-06    &   0.41    &   -1.52   &   3.59    &   -0.54   $^{+    0.26    }_{-    0.2 }$  &   -4.64   $^{+    2.3 }_{-    100 }$  &   64  $^{+    10  }_{-    20  }$  &   1.15    &   70.62   &   26   \\
010222  &   72  &   1.87E-05    &   5.25E-05    &   0.36    &   -1.57   &   2.05    &   -1.59   $^{+    0.01    }_{-    0.01    }$  &   -2.20  &   10000   $^{+    100 }_{-    9200    }$  &   1.39    &   21.92   &   27   \\
010304  &   24  &   1.45E-06    &   3.67E-06    &   0.39    &   -1.47   &   3.42    &   -0.45   $^{+    0.53    }_{-    0.26    }$  &   -2.11   $^{+    0.3 }_{-    0.32    }$  &   43  $^{+    18  }_{-    22  }$  &   0.71    &   77.3    &   26   \\
010320  &   31  &   4.58E-07    &   2.18E-10    &   2096.95 &   -2.24   &   2.84    &   3.00    $^{+    100 }_{-    3   }$  &   -10 $^{+    6.9 }_{-    100 }$  &   9   $^{+    1   }_{-    1   }$  &   0.77    &   61.57   &   27   \\
010324  &   370 &   5.74E-06    &   9.25E-06    &   0.62    &   -1.74   &   1.18    &   -1.53   $^{+    0.47    }_{-    0.25    }$  &   -2.20   &   90  $^{+    10000   }_{-    55  }$  &   1.15    &   2.02    &   28   \\
010412  &   102 &   4.87E-06    &   2.45E-05    &   0.20    &   -1.36   &   2.01    &   -1.18   $^{+    0.06    }_{-    0.06    }$  &   -8.63   $^{+    6.4 }_{-    100 }$  &   283 $^{+    91  }_{-    52  }$  &   0.8 &   36.69   &   27   \\
010501A &   78  &   3.87E-07    &   7.53E-07    &   0.51    &   -1.62   &   1.28    &   -1.21   $^{+    0.33    }_{-    0.27    }$  &   -2.20   &   67  $^{+    110 }_{-    33  }$  &   1.09    &   6.41    &   27   \\
010501B &   61  &   7.06E-07    &   2.17E-06    &   0.33    &   -1.33   &   1.66    &   0.31    $^{+    1.23    }_{-    0.62    }$  &   -2.73   $^{+    0.8 }_{-    100 }$  &   50  $^{+    27  }_{-    27  }$  &   0.6 &   31.94   &   27   \\
010518  &   47  &   4.50E-07    &   1.26E-06    &   0.36    &   -1.53   &   1.19    &   -1.01   $^{+    0.62    }_{-    0.42    }$  &   -2.20  &   68  $^{+    212 }_{-    35  }$  &   1.04    &   5.24    &   27   \\
010527  &   75  &   8.60E-07    &   6.05E-07    &   1.42    &   -2.06   &   1.37    &   -1.00   &   -2.22   $^{+    0.2 }_{-    100 }$  &   10  $^{+    27  }_{-    10  }$  &   1.33    &   2.45    &   27   \\
010707  &   20  &   1.98E-07    &   5.46E-10    &   362.39  &   -2.14   &   2.27    &   -0.11   $^{+    1.07    }_{-    0.9 }$  &   -7.45   $^{+    4.9 }_{-    100 }$  &   7   $^{+    1   }_{-    1   }$  &   1.5 &   25.33   &   27   \\
011030  &   310 &   7.36E-07    &   2.46E-07    &   3.00    &   -1.96   &   1.28    &   -1.00   &   -2.6    $^{+    0.6 }_{-    100 }$  &   8   $^{+    3   }_{-    2   }$  &   1.05    &   7.72    &   28   \\
011121  &   185 &   1.58E-05    &   6.77E-05    &   0.23    &   -1.42   &   2.89    &   -1.44   $^{+    0.06    }_{-    0.01    }$  &   -2.20   &   10000   $^{+    100 }_{-    9300    }$  &   0.83    &   60.57   &   28   \\
011211  &   275 &   2.04E-06    &   1.51E-06    &   1.35    &   -1.9    &   2.99    &   -0.89   $^{+    0.46    }_{-    0.38    }$  &   -2.3    $^{+    0.2 }_{-    0.52    }$  &   14  $^{+    15  }_{-    4   }$  &   0.93    &   61.6    &   27   \\
020113  & 320 & 6.783E-6 & 1.231E-6 & 0.55 & -1.49 & 2.05 & -1.0$\pm$1.3 & -1.8$\pm$1.3 & 12$^{+ 40 }_{- 2}$ & 2.16 & -2.97 & 27 \\
020118  &   50  &   1.20E-06    &   6.53E-07    &   1.84    &   -2.08   &   1.58    &   -0.98   $^{+    0.86    }_{-    0.36    }$  &   -4.08   $^{+    3.1 }_{-    100 }$  &   26  $^{+    8   }_{-    26  }$  &   0.57    &   29.42   &   26   \\
020321  &   57  &   2.64E-07    &   1.54E-06    &   0.17    &   -1.31   &   0.96    &   -1.18   $^{+    0.1 }_{-    0.1 }$  &   -2.20  &   421 $^{+    680 }_{-    170 }$  &   0.98    &   5.60 &   28   \\
020322  &   58  &   5.39E-07    &   6.24E-07    &   0.86    &   -1.84   &   1.05    &   -1.00   &   -2.06   $^{+    0.2 }_{-    0.71    }$  &   17  $^{+    21  }_{-    8   }$  &   0.89    &   5.37    &   27   \\
020409  &   60  &   9.59E-07    &   1.38E-06    &   0.70    &   -1.78   &   0.93    &   -1.31   $^{+    0.68    }_{-    0.43    }$  &   -2.20  &   41  $^{+    200 }_{-    21  }$  &   0.82    &   4.01    &   28   \\
020427  &   56  &   5.61E-07    &   3.10E-07    &   1.81    &   -2.20   &   0.38    &   -1.00   &   -2.20   $^{+    0.1 }_{-    0.16    }$  &   1   $^{+    4   }_{-    1   }$  &   0.39    &   0.11    &   27   \\
    \noalign{\smallskip}
    \hline
\end{longtable}

\clearpage
\twocolumn

\begin{table*}
\caption[]{Average and width of the parent distributions of spectral parameters.}
\label{tab:spectralaverage}
\centering                         
\begin{tabular}{cccccccc}      
\hline\hline               
    Class & $<log(\epc)>$ & $\rm \sigma_{log(\epc)}$ & $<\alpha>$ & $\sigma_{\alpha}$ & $<\beta>$ & $\sigma_{\beta}$\\
    \noalign{\smallskip}
    \hline
    \noalign{\smallskip}
XRF &   $0.93\pm0.05$      &    $0.37\pm0.10$  &    $-1.35\pm0.23$   &    $0.10\pm0.10$ &   $-2.09\pm0.12$   &  $0.16\pm0.06$\\
XRR &   $1.53\pm0.14$     &    $0.25\pm0.08$   &   $-0.98\pm0.17$ &  $0.26\pm0.11$  &    $-1.83\pm0.09$   &   $0.14\pm0.05$ \\
GRB &   $1.92\pm0.21$   &   $0.40\pm0.12$   &   $-0.96\pm0.16$ &   $0.33\pm0.10$    &   $-1.41\pm0.12$ &   $0.13\pm0.07$  \\
ALL &   $1.38\pm0.11$   &    $0.42\pm0.08$    &    $-0.99\pm0.11$    &   $0.31\pm0.08$ &   $-1.89\pm0.10$ &   $0.26\pm0.06$ \\
    \hline     
\end{tabular}
\tablefoot{The first column indicate the sample class. In the 2nd, 4th, and 6th columns the average of the parent distribution is reported for $log(\epc)$, $\alpha$, and $\beta$ respectively. While in the 3nd, 5th, and 7th columns the width of the parent distribution is reported for $log(\epc)$, $\alpha$, and $\beta$ respectively. } 
\end{table*}

\section{Spectra}
\label{App:spectra}
\begin{figure}
 \includegraphics[width=0.5\textwidth]{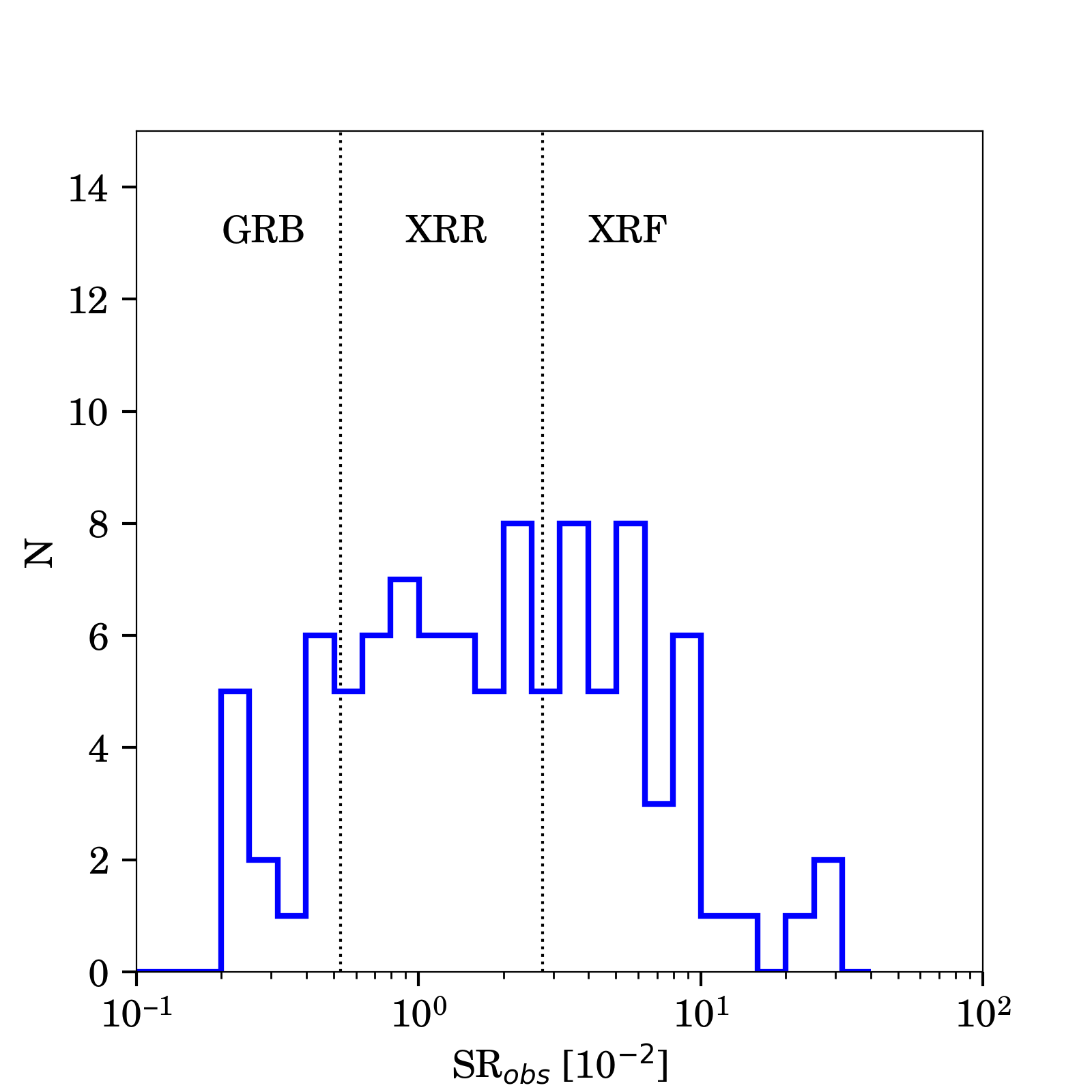}
\caption{Distribution of the {\it observed} softness ratio $\srobs$, given by the ratio of the WFC (2-28 keV) and GRBM (40-700 keV) count
rates.  GRBs have $\srobs \le 0.53\times 10^{-2}$; XRRs 
$0.53\times 10^{-2} \le \srobs \le 2.75\times 10^{-2}$; XRFs $\srobs \ge 2.75\times 10^{-2}$.}
\label{fig:SR_obs}
\end{figure}

In this Appendix we include some examples of spectral fits. Each spectrum was fitted with a black body, a power law and a Band model. We also include the $\rm SR_{obs}$ distribution in Fig.\ref{fig:SR_obs}.

\begin{figure}
 \includegraphics[width=0.5\textwidth]{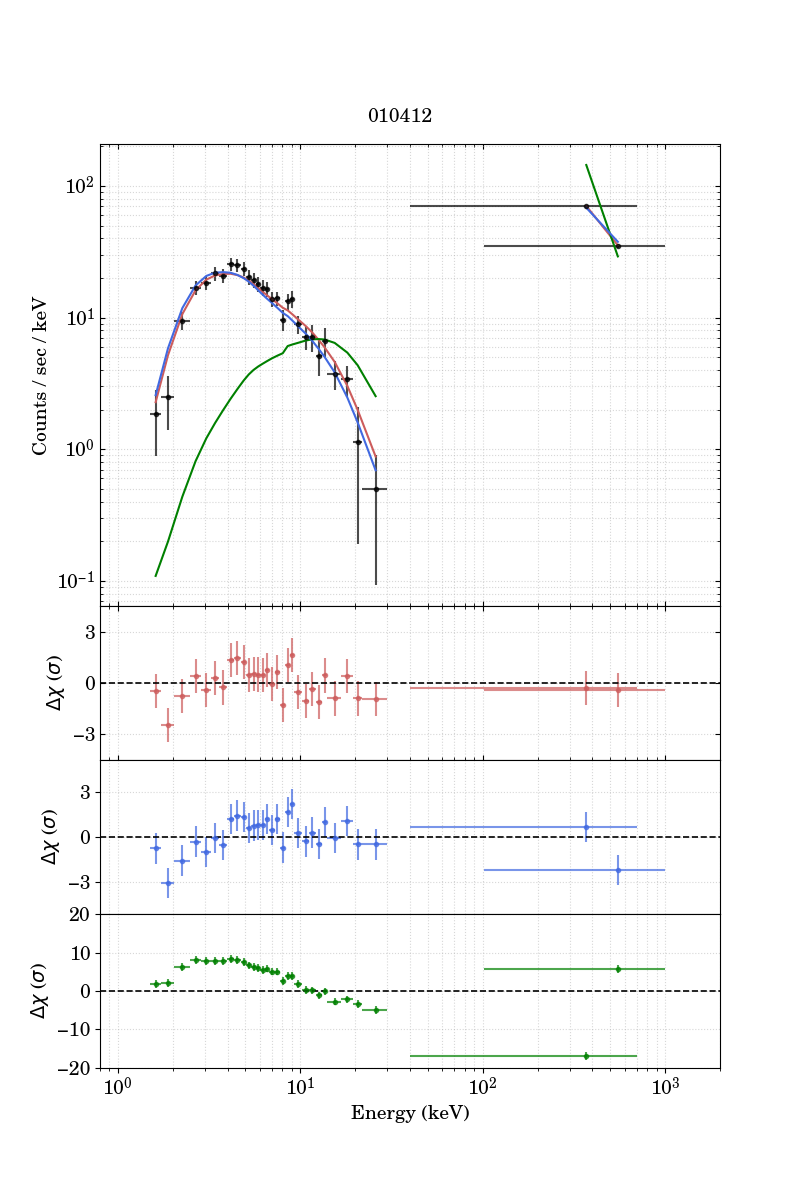}
\caption{Top panel: WFC and GRBM spectrum of GRB 010412. The three fitted models are represented with three lines: in red the Band model, in blue the power law model, and in green the black body. In the panels below, residuals are represented for each model, with the same color code.}
\label{fig:GRB_spectrum}
\end{figure}

\begin{figure}
 \includegraphics[width=0.5\textwidth]{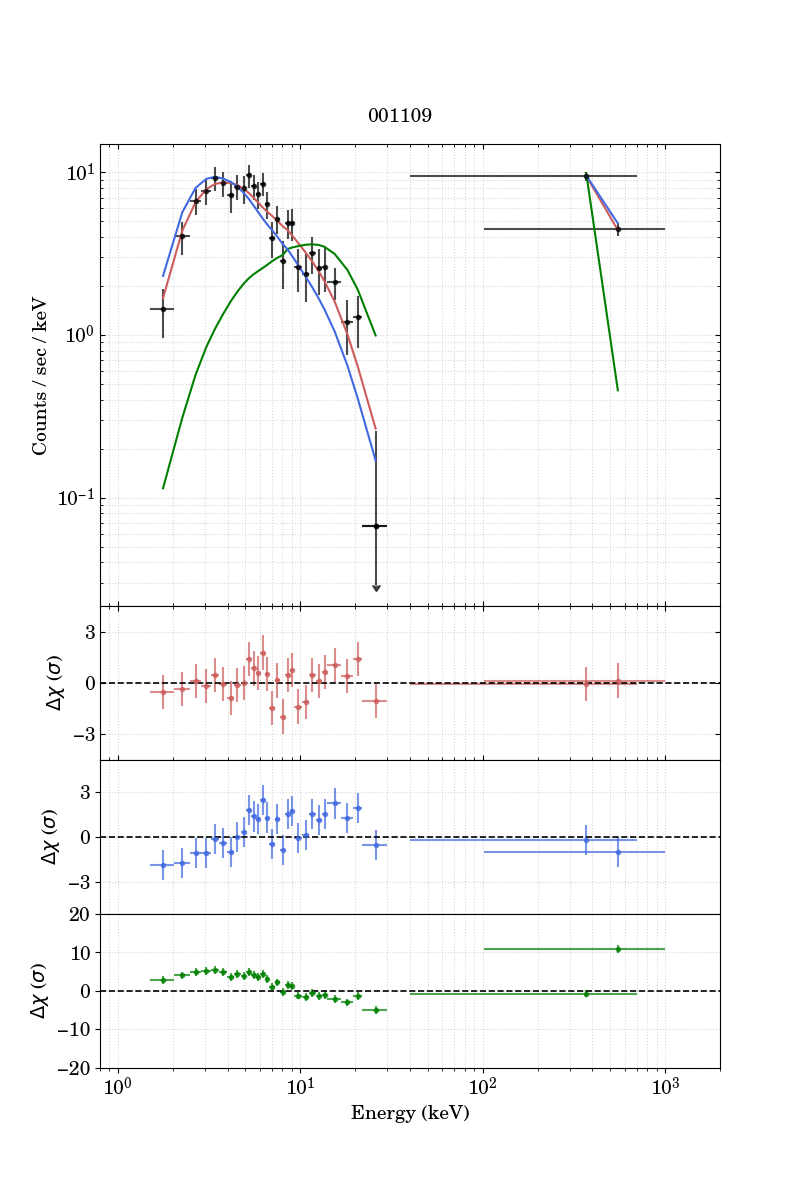}
\caption{Same as Fig.\ref{fig:GRB_spectrum}, but for XRR 001109.}
\label{fig:XRR_spectrum}
\end{figure}

\begin{figure}
 \includegraphics[width=0.5\textwidth]{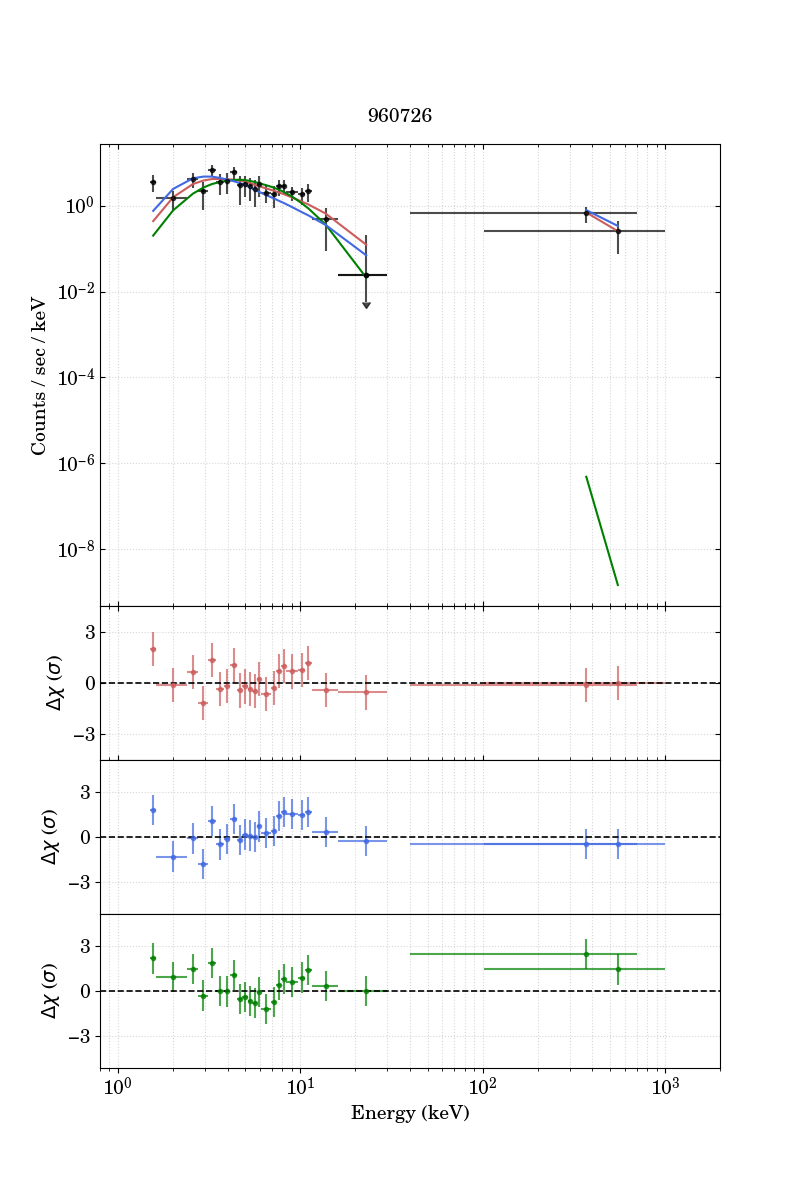}
\caption{Same as Fig.\ref{fig:GRB_spectrum}, but for XRF 960726.}
\label{fig:XRF_spectrum}
\end{figure}

GRB 010402 is represented in Fig.\ref{fig:GRB_spectrum}, for the $\cq$ values for the power law and Band model, we refer the reader to Table \ref{tab:spectra}, while the black body (BB) $\cq=1279.02$. 
An example of XRR (001109) is represented in Fig.\ref{fig:XRR_spectrum}. For the BB model, $\cq=439.51$.
Finally, an example of XRF spectrum (960726) is in Fig. \ref{fig:XRF_spectrum}. In this case, the BB model has $\cq=43.14$. 
We note that the BB mode fits, having a narrow spectral shape, are strongly influenced by the statistical weight of the GRBM data points. A highly significant GRBM detection (the GRB case) forces the BB peak to be positioned between the GRBM and WFC energy bands. When the GRBM data points have low significance (the XRF case), the fit statistic is dominated by the WFC data, shifting the BB peak fully into the WFC band. The XRR case represents an intermediate scenario, where the BB peak is located at lower energies compared to the GRB case, yet it still fails to provide a good fit to the WFC data.
In all cases, the preferred model is the Band one. 

We also note that for the objects in our sample the WFC are not sensitive to intrinsic  column densities as high as about 10$^{23} \rm cm^{-2}$ at $z\simeq2$ (the average redshift of GRBs) which is well above the average value observed in GRB afterglows (0.05 $\times 10^{23} \rm cm^{-2}$; \citealt{Campana2014}). Furthermore, as described in the Section \ref{subsec:spec_shape}, the Band model provides an adequate model of the full population and does not require additional components.
As a final confirmation we have verified whether fits with reduced $\cnu \geq 1.5$ could improve with an additional column density added to the Band model. We do not find any significant improvement and any additional column density is consistent with 0.

\end{appendix}

\end{document}